\documentclass[reqno]{amsart}
\usepackage{amsmath,amsthm,amssymb,eucal,mathrsfs,fullpage,setspace}
\usepackage{graphicx}
\usepackage[all,cmtip]{xy}
\usepackage[round]{natbib}
\usepackage{lineno}
\usepackage{subfig}

\newcommand{\fig}[1]{Fig.~\ref{fig:#1}}
\newcommand{\eq}[1]{Eq.~(\ref{eq:#1})}
\newcommand{\thm}[1]{Theorem~\ref{thm:#1}}
\newcommand{\defn}[1]{Definition~\ref{def:#1}}
\newcommand{\cor}[1]{Corollary~\ref{cor:#1}}

\theoremstyle{definition}
\newtheorem{corollary}{Corollary}
\newtheorem{definition}{Definition}
\newtheorem{example}{Example}

\newtheorem{theorem}{Theorem}
\newtheorem*{repeatedtheorem}{\thm{maintheorem}}
\newtheorem*{repeateddefinition}{\defn{spatiallyadditive}}
\newtheorem*{repeatedcorollary}{\cor{spatiallyadditive}}

\title{Asymmetric evolutionary games}

\author{Alex McAvoy and Christoph Hauert}

\begin{document}

\begin{abstract}
Evolutionary game theory is a powerful framework for studying evolution in populations of interacting individuals. A common assumption in evolutionary game theory is that interactions are symmetric, which means that the players are distinguished by only their strategies. In nature, however, the microscopic interactions between players are nearly always asymmetric due to environmental effects, differing baseline characteristics, and other possible sources of heterogeneity. To model these phenomena, we introduce into evolutionary game theory two broad classes of asymmetric interactions: ecological and genotypic. Ecological asymmetry results from variation in the environments of the players, while genotypic asymmetry is a consequence of the players having differing baseline genotypes. We develop a theory of these forms of asymmetry for games in structured populations and use the classical social dilemmas, the Prisoner's Dilemma and the Snowdrift Game, for illustrations. Interestingly, asymmetric games reveal essential differences between models of genetic evolution based on reproduction and models of cultural evolution based on imitation that are not apparent in symmetric games.
\end{abstract}

\maketitle

%

\section{Introduction}

Evolutionary game theory has been used extensively to study the evolution of cooperation in social dilemmas \citep{ohtsuki:Nature:2006,nowak:Science:2006,taylor:Nature:2007}. A social dilemma is typically modeled as a game with two strategies, cooperate ($C$) and defect ($D$), whose payoffs for pairwise interactions are defined by a matrix of the form
\begin{linenomath}
\begin{align}\label{generic2by2}
\bordermatrix{%
& C & D \cr
C &\ R, R & \ S, T \cr
D &\ T, S & \ P, P \cr
}
\end{align}
\end{linenomath}
\citep{maynardsmith:CUP:1982,hofbauer:CUP:1998}. For a focal player using a strategy on the left-hand side of this matrix against an opponent using a strategy on the top of the matrix, the first (resp. second) coordinate of the corresponding entry of this matrix is the payoff to the focal player (resp. opponent). That is, a cooperator receives $R$ when facing another cooperator and $S$ when facing a defector; a defector receives $T$ when facing a cooperator and $P$ when facing another defector. Since the same argument applies to the opponent, the game defined by (\ref{generic2by2}) is \textit{symmetric}. If defection pays more than cooperation when the opponent is a cooperator ($T>R$), but the payoff for mutual cooperation is greater than the payoff for mutual defection ($R>P$), then a social dilemma \citep{dawes:ARP:1980,hauert:JTB:2006a} arises from this game due to the conflict of interest between the individual and the group (or pair). The nature of this social dilemma depends on the ordering of $R$, $S$, $T$, and $P$. Biologically, the most important rankings are given by the Prisoner's Dilemma ($T>R>P>S$) and the Snowdrift Game ($T>R>S>P$) \citep{maynardsmith:CUP:1982,hauert:Nature:2004,doebeli:EL:2005,hauert:JTB:2006a,voelkl:Games:2010}.

Since matrix ({\ref{generic2by2}}) defines a symmetric game, any two players using the same strategy are indistinguishable for the purpose of calculating payoffs. In nature, however, asymmetry frequently arises in interspecies interactions such as parasitic or symbiotic relationships \citep{maynardsmith:CUP:1982}. Interactions between subpopulations, such as in Dawkins' Battle of the Sexes Game \citep{dawkins:OUP:1976,schuster:AB:1981,smith:TPB:1987,hofbauer:JMB:1996}, also give rise to asymmetry that cannot be modeled by the symmetric matrix (\ref{generic2by2}). Even intraspecies interactions are essentially always asymmetric: (i) phenotypic variations such as size, strength, speed, wealth, or intellectual capabilities; (ii) differences in access to and availability of environmental resources; or (iii) each individual's history of past interactions, all affect the interacting individuals differently and result in asymmetric payoffs. The winner-loser effect, for example, is a well-studied example of effects of previous encounters on future interactions and has been reported across taxa \citep{dugatkin:BE:1997,maynardsmith:CUP:1982}, including even mollusks \citep{wright:JEMBE:1993,shanks:BE:2002}. Asymmetry may also result from the assignment of social roles \citep{selten:JTB:1980,hammerstein:AB:1981,ohtsuki:JTB:2010}, such as the roles of ``parent" and ``offspring" \citep{marshall:JTB:2009}: cooperation may be tied to individual energy or strength, for example, which is, in turn, determined by a player's role. In the realm of continuous strategies, adaptive dynamics has been used to study asymmetric competition, which applies to the resource consumption of plants, for instance \citep{weiner:TEE:1990,freckleton:FE:2001,doebeli:JMB:2012}. In social dilemmas containing many cooperators, accumulated benefits may be synergistically enhanced (or discounted) in a way that depends on who or where the players are \citep{hauert:JTB:2006a}, thereby making larger group interactions asymmetric. To model such interactions using evolutionary game theory, the payoff matrix must reflect the asymmetry.

In the \textit{Donation Game}, a cooperator pays a cost, $c$, to deliver a benefit, $b$, to the opponent, while a defector pays no cost and provides no benefit \citep{sigmund:PUP:2010}. In terms of matrix (\ref{generic2by2}), this game satisfies $R=b-c$, $S=-c$, $T=b$, and $P=0$. Provided $b$ and $c$ are positive, mutual defection is the only Nash equilibrium. If $b>c$, then this game defines a Prisoner's Dilemma. Perhaps the simplest way to modify this game to account for possible sources of asymmetry is to allow for each pair of players to have a distinct payoff matrix; that is, the payoff matrix for player $i$ against player $j$ in the Donation Game is
\begin{linenomath}
\begin{align}\label{donationmatrix}
\mathbf{M}^{ij} := \bordermatrix{%
 & C & D \cr
 C &\ b_{j}-c_{i},\ b_{i}-c_{j} & \ -c_{i},\ b_{i} \cr
 D &\ b_{j},\ -c_{j} & \ 0,\ 0 \cr
}
\end{align}
\end{linenomath}
for some $b_{i}$, $b_{j}$, $c_{i}$, and $c_{j}$. If player $i$ cooperates, then this player donates $b_{i}$ to his or her opponent and incurs a cost of $c_{i}$ for doing so. As before, defectors provide no benefit and pay no cost. The index $i$ could refer to a baseline trait of the player, the player's location, his or her history of past interactions, motivation \citep{bergman:PRSB:2010}, or any other non-strategy characteristic that distinguishes one player from another.

Games based on matrices of the form (\ref{donationmatrix}), with payoffs for \textit{both} players in each entry of the matrix, are sometimes called \textit{bimatrix} games. Although bimatrix games have appeared in the context of evolutionary dynamics \citep{hofbauer:JMB:1996,hofbauer:BAMS:2003,ohtsuki:JTB:2010}, most of the focus on these games has been in the setting of classical game theory and economics \citep[see][]{fudenberg:MIT:1991} where ``matrix game" generally means ``bimatrix game." Bimatrix games may be used to model classical asymmetric interactions such as those arising from sexual asymmetry in the Battle of the Sexes Game \citep{magurran:PRSB:1991}. The asymmetric, four-strategy Hawk-Dove Game of \citep{maynardsmith:CUP:1982} consisting of the strategies Hawk, Dove, Bourgeois, and anti-Bourgeois may also be framed as a ($4\times 4$) bimatrix game \citep[see][]{mesterton_gibbons:EE:1992}. Symmetric matrix games, such as (\ref{generic2by2}), are special cases of bimatrix games. We explore here the ways in which bimatrix games can be incorporated into evolutionary dynamics and used to model natural asymmetries in biological populations.

We treat two particular forms of asymmetry: \textit{ecological} and \textit{genotypic}. Ecological asymmetry is derived from the locations of the players, whereas genotypic asymmetry is based on the players themselves. With ecological asymmetry, $\mathbf{M}^{ij}$ is the payoff matrix for a player at \textit{location} $i$ against a player at \textit{location} $j$. Since the payoffs depend on the locations of the players, this form of asymmetry requires a structured population. Ecological asymmetry is a natural consideration in evolutionary dynamics since it ties strategy success to the environment. In the Donation Game, for instance, cooperators might be donating goods or services, but the costs and benefits may depend on the environmental conditions, i.e. the location of the donor.

On the other hand, players might instead differ in ability or strength, and ``strong" cooperators might contribute greater benefits (or incur lower costs) than ``weak" cooperators. This variation results in genotypic asymmetry, where each player has a baseline genotype (strength) and a strategy ($C$ or $D$). This form of asymmetry turns out to be subtler than it seems at first glance, however, since genotypes are generally represented by strategies in evolutionary game theory \citep{maynardsmith:CUP:1982,dugatkin:OUP:2000}. In particular, it might seem that the genotype and strategy of a player could be combined into a single composite strategy and that the symmetric game based on these composite strategies could replace the original asymmetric game. As it happens, whether genotypic asymmetry can be resolved by a symmetric game depends on the details of the evolutionary process.

Classically, evolutionary games were studied in infinite populations via replicator dynamics \citep{taylor:MB:1978}, and more recently these games have been considered in finite populations \citep{nowak:Nature:2004,taylor:BMB:2004}. Because every biological population is finite, we focus on finite populations (which, for technical reasons, we assume to be large). Since ecological asymmetry requires distinguishing different locations within the population, we assume that the population is structured and that a network defines the structure. Network-structured populations have received a considerable amount of attention in evolutionary game theory and provide a natural setting in which to study social dilemmas \citep{lieberman:Nature:2005,ohtsuki:Nature:2006,ohtsuki:JTB:2006,taylor:Nature:2007,szabo:PR:2007,debarre:NC:2014}. Compared to well-mixed populations, in which each player interacts with every other player, networks can restrict the interactions that occur within the population by specifying which players are ``neighbors,'' i.e. share a link. We represent the links among the $N$ players in the population using an adjacency matrix, $\left(w_{ij}\right)_{1\leqslant i,j\leqslant N}$, which is defined by letting $w_{ij}=1$ if there is a link from vertex $i$ to vertex $j$ and $0$ otherwise (and satisfies $w_{ij}=w_{ji}$ for each $i$ and $j$).

In an evolutionary game, the \textit{state} of a population of players is defined by specifying the strategy of each player. Each player interacts with all of his or her neighbors. The total payoff to a player is multiplied by a selection intensity, $\beta\geqslant 0$, and then converted into \textit{fitness} (see Methods). Once each player is assigned a fitness, an update rule is used to determine the state of the population at the next time step \citep{nowak:BP:2006}. For example, with a birth-death update rule, a player is chosen from the population for reproduction with probability proportional to relative fitness. A neighbor of the reproducing player is then randomly chosen for death, and the offspring, who inherits the strategy of the parent, fills the vacancy. This process is a modification of the Moran process \citep{moran:MPCPS:1958}, adapted to allow for (i) frequency-dependent fitnesses and (ii) population structures that are not necessarily well-mixed. The order of birth and death could also be reversed to get a death-birth update rule \citep{ohtsuki:Nature:2006}. In this rule, death occurs at random and the neighbors of the deceased compete to reproduce in order to fill the vacancy. These two rules result in the update of a single strategy in each time step, but one could consider other rules, such as Wright-Fisher updating, in which \textit{all} of the strategies are revised in each generation \citep{imhof:JMB:2006}. The rules mentioned to this point define strategy updates via reproduction and inheritance; as such, we refer to them as \textit{genetic} update rules.

Another popular class of update rules is based on revisions to the existing players' strategy choices. We refer to rules falling into this class as \textit{cultural} update rules. Examples include imitation updating, in which a player is selected at random to evaluate his or her strategy and then probabilistically compares this strategy to those of his or her neighbors \citep{ohtsuki:Nature:2006}. A more localized version of this update rule is known as pairwise comparison updating, in which a player chooses a random neighbor for comparison rather than looking at the entire neighborhood \citep{szabo:PRE:1998,traulsen:JTB:2007}. Under \textit{best response dynamics}, an individual adopts the strategy that performs best given the current strategies of his or her neighbors \citep{ellison:E:1993}. In each of these cultural processes, the strategy of a player can change, but the underlying genotype is always the same, which suggests that baseline genotype and strategy need to be treated separately.

Genotypic asymmetry needs to be handled more carefully if the update rule is genetic since the nature of genotype transmission affects the dynamics of the process. In contrast to cultural processes, the genotype \textit{and} strategy of a player at a given location may both change if the update rule is genetic: genotype may be inherited but not imitated. We will see that this property results in cultural and genetic processes behaving completely differently in the presence of genotypic asymmetry.  \textit{Phenotype} may have both genetic and environmental components \citep{mahner:JTB:1997,baye:PM:2011}, and after treating the genetic (genotypic) and environmental components separately, these two forms of asymmetry may be combined in order to get a model in which the asymmetry is derived from varying baseline phenotypes. Thus, with a theory of both ecological asymmetry and genotypic asymmetry based on inherited genotypes, one can account for more complicated forms of asymmetry appearing in biological populations.

\section{Results}

\subsection{Ecological asymmetry}

Here we develop a framework for \textit{ecologically} asymmetric games in which the payoffs depend on the locations of the players as well as their strategies. We assume that all of the players have the same set of strategies (or ``actions") available to them, $\left\{A_{1},\dots ,A_{n}\right\}$. The payoff matrix for a player at vertex $i$ against a player at vertex $j$ is
\begin{linenomath}
\begin{align}\label{eq:ecologicalmatrix}
\mathbf{M}^{ij} = \bordermatrix{%
 & A_{1} & A_{2} & \cdots & A_{n} \cr
A_{1} &\ a_{11}^{ij}, a_{11}^{ji} & \ a_{12}^{ij}, a_{21}^{ji} & \ \cdots & \ a_{1n}^{ij}, a_{n1}^{ji} \cr
A_{2} &\ a_{21}^{ij}, a_{12}^{ji} & \ a_{22}^{ij}, a_{22}^{ji} & \ \cdots & \ a_{2n}^{ij}, a_{n2}^{ji} \cr
\ \vdots &\ \vdots & \ \vdots & \ \ddots & \ \vdots \cr
A_{n} &\ a_{n1}^{ij}, a_{1n}^{ji} & \ a_{n2}^{ij}, a_{2n}^{ji} & \ \cdots & \ a_{nn}^{ij}, a_{nn}^{ji} \cr
} .
\end{align}
\end{linenomath}
That is, a player at vertex $i$ using strategy $A_{r}$ against an opponent at vertex $j$ using strategy $A_{s}$ realizes a payoff of $a_{rs}^{ij}$, whereas his opponent receives $a_{sr}^{ji}$. Since $a_{rs}^{ij}$ depends on $i$ and $j$, these payoff matrices capture the asymmetry of the game.

In the simpler setting of symmetric games, the pair approximation method has been used successfully to describe the dynamics of evolutionary processes on networks \citep{matsuda:PTP:1992,bollobas:CUP:2001,ohtsuki:Nature:2006,vukov:PRE:2006,ohtsuki:JTB:2006}. For each $r\in\left\{1,\dots ,n\right\}$, this method approximates the frequency of strategy $A_{r}$, which we denote by $p_{r}$, using the frequencies of strategy pairs in the population. Pair approximation is expected to be accurate on large \textit{random regular} networks \citep{bollobas:CUP:2001,ohtsuki:Nature:2006}, so we assume that the network is regular (of degree $k>2$) and that $N$ is sufficiently large. (For $k=2$, the network is just a cycle, which we do not treat here.) We also take $\beta\ll 1$, meaning that selection is \textit{weak}, which results in a separation of timescales: the local configurations equilibrate quickly, while the global strategy frequencies change much more slowly. This separation allows us to get an explicit expression for the expected change, $\mathbb{E}\left[\Delta p_{r}\right]$, in the frequency of strategy $A_{r}$ for each $r$. Incidentally, weak selection happens to be quite reasonable from a biological perspective since each trait is expected to have only a small effect on the overall fitness of a player \citep{wu:PRE:2010,tarnita:PNAS:2011,wu:PLoSCB:2013}.

Interestingly, for two genetic and two cultural update rules, weak selection reduces ecological asymmetry to a symmetric game derived from the spatial average of the payoff matrices:

\begin{theorem}\label{thm:maintheorem}
In the limit of weak selection, the dynamics of the ecologically asymmetric death-birth, birth-death, imitation, and pairwise comparison processes on a large, regular network may be approximated by the dynamics of a \textit{symmetric} game with the same update rule and payoff matrix $\overline{\mathbf{M}}:=\frac{1}{kN}\sum_{i,j=1}^{N}w_{ij}\mathbf{M}^{ij}$, i.e.
\begin{linenomath}
\begin{align}
\overline{\mathbf{M}} &= \bordermatrix{%
 & A_{1} & A_{2} & \cdots & A_{n} \cr
A_{1} &\ \overline{a}_{11}, \overline{a}_{11} & \ \overline{a}_{12}, \overline{a}_{21} & \ \cdots & \ \overline{a}_{1n}, \overline{a}_{n1} \cr
A_{2} &\ \overline{a}_{21}, \overline{a}_{12} & \ \overline{a}_{22}, \overline{a}_{22} & \ \cdots & \ \overline{a}_{2n}, \overline{a}_{n2} \cr
\ \vdots &\ \vdots & \ \vdots & \ \ddots & \ \vdots \cr
A_{n} &\ \overline{a}_{n1}, \overline{a}_{1n} & \ \overline{a}_{n2}, \overline{a}_{2n} & \ \cdots & \ \overline{a}_{nn}, \overline{a}_{nn} \cr
} ,
\end{align}
\end{linenomath}
where $\overline{a}_{st}:=\frac{1}{kN}\sum_{i,j=1}^{N}w_{ij}a_{st}^{ij}$ for each $s$ and $t$.
\end{theorem}
For a proof of \thm{maintheorem}, see Methods. In Methods, we derive explicit formulas for $\mathbb{E}\left[\Delta p_{r}\right]$ for each $r$ (where $p_{r}$ is the frequency of strategy $A_{r}$ and $\mathbb{E}\left[\Delta p_{r}\right]$ is the expected change in $p_{r}$ in one step of the process) and show that these expectations depend on $\overline{\mathbf{M}}$ in the limit of weak selection. If we choose an appropriate time scale and make the approximation
\begin{linenomath}
\begin{align}
\dot{p}_{r} &:= \frac{dp_{r}}{dt} = \frac{\mathbb{E}\left[\Delta p_{r}\right]}{\Delta t} ,
\end{align}
\end{linenomath}
then the dynamics of an ecologically asymmetric process may also be described in terms of the replicator equation (on graphs) of \cite{ohtsuki:JTB:2006}: If $\phi :=\sum_{s,t=1}^{n}p_{s}p_{t}\overline{a}_{st}$, then
\begin{linenomath}
\begin{align}
\dot{p}_{r} &= p_{r}\left(\sum_{s=1}^{n}p_{s}\left(\overline{a}_{rs}+\overline{b}_{rs}\right) - \phi\right) ,
\end{align}
\end{linenomath}
where $\overline{b}_{rs}$ is a function of $\overline{\mathbf{M}}$, $k$, and the update rule. (For each of the four processes, the explicit expression for $\overline{b}_{rs}$ is provided in Methods.) The matrix $\left(\overline{b}_{rs}\right)_{r,s=1}^{n}$ accounts for local competition resulting from the population structure \citep[see][]{ohtsuki:JTB:2006}. In particular, the Ohtsuki-Nowak transform,
\begin{linenomath}
\begin{align}
\left(\overline{a}_{rs}\right)_{r,s=1}^{n}\longrightarrow\left(\overline{a}_{rs}+\overline{b}_{rs}\right)_{r,s=1}^{n},
\end{align}
\end{linenomath}
which transforms the classical replicator equation into the replicator equation on graphs, also applies to evolutionary games with ecological asymmetry.

Even though interactions are now governed by a symmetric game, \thm{maintheorem} states that, in general, the dynamics depend on the particular network configuration, $\left(w_{ij}\right)_{1\leqslant i,j\leqslant N}$; that is, the symmetric payoffs defined by $\overline{\mathbf{M}}$ still depend on the network structure, or, equivalently, on the distribution of ecological resources within the population. However, somewhat surprisingly, there is a broad class of games for which this dependence vanishes:
\begin{definition}\label{def:spatiallyadditive}
If $a_{rs}^{ij}=x_{rs}^{i}+y_{rs}^{j}$ for each $r$ and $s$, then $\mathbf{M}^{ij}$ is called a \textit{spatially additive} payoff matrix. If $\mathbf{M}^{ij}$ is spatially additive for each $i$ and $j$, then the game is said to be spatially additive.
\end{definition}

A game is spatially additive if the payoff for an interaction between any two members of the population can be decomposed as a sum of two components, one from each player's location. Note that spatial additivity is different from the ``equal gains from switching" property \citep{nowak:AAM:1990} in that neither implies the other. However, spatial additivity is an analogue in the following sense: if two players at different locations use the same strategy against a common opponent, then the difference in these two players' payoffs for this interaction is independent of the location of the opponent. Interchanging ``location" and ``strategy," one obtains the equal gains from switching property. The importance of spatially additive games is due to the following corollary to \thm{maintheorem}:
\begin{corollary}\label{cor:spatiallyadditive}
If $\mathbf{M}^{ij}$ is spatially additive for each $i$ and $j$, then the expected change in the frequency of strategy $A_{r}$, $\mathbb{E}\left[\Delta p_{r}\right]$, is independent of $\left(w_{ij}\right)_{1\leqslant i,j\leqslant N}$ for each $r$. In particular, the dynamics of the process do not depend on the particular network configuration.
\end{corollary}

\begin{figure*}
\begin{center}
\includegraphics[width=1.0\textwidth]{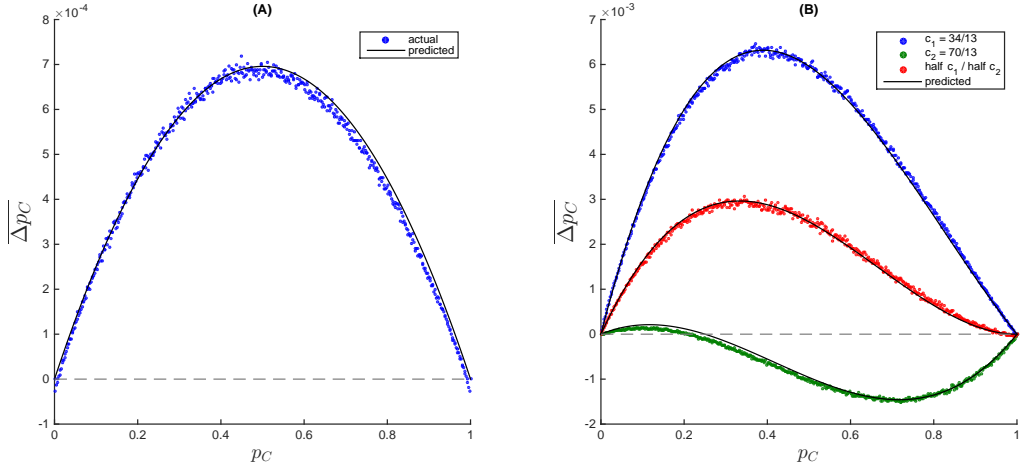}
\end{center}
\caption{Average change in the frequency of cooperators, $\overline{\Delta p_{C}}$, as a function of the frequency of cooperators, $p_{C}$, in (A) an asymmetric Donation Game and (B) asymmetric Snowdrift Games. The update rules are (A) imitation and (B) death-birth, and each process has for a selection intensity $\beta =0.01$. In both figures, the network is a random regular graph of size $N=500$ and degree $k=3$. In (A), benefits and costs of cooperation vary across vertices according to a Gaussian distribution with mean $3.5$, variance $1.0$ for benefits and mean $0.5$, variance $0.25$ for costs. In (B), the benefit is $b=5.0$ for all vertices, and the costs are either low, $c_1=34/13$, or high $c_2=70/13$, which actually recovers the payoff ranking of the Prisoner's Dilemma because $c_2>b$. The costs are the same for all vertices ($c_1$, blue and $c_2$, green) or mixed at equal proportions (red). (B) confirms that the average change in cooperators in the mixed Snowdrift Game/Prisoner's Dilemma (red) may be obtained by averaging these changes for the Snowdrift Game (blue) and the Prisoner's Dilemma (green). The small, systematic deviations between simulation data and analytical predictions (solid lines) are explained in Methods (where it is also shown that $\overline{\Delta p_{C}}$ is linear in $\beta$ for $\beta\ll 1$).\label{fig:Ecological}}
\end{figure*}

As an example, the asymmetric Donation Game is spatially additive \textit{and} possesses the equal gains from switching property, which greatly simplifies the analysis of its dynamics:

\begin{example}[Donation Game with ecological asymmetry]\label{ex:donationEcological}
The asymmetric Donation Game with payoff matrices defined by Eq. (\ref{donationmatrix}) is spatially additive and satisfies
\begin{linenomath}
\begin{align}
\overline{\mathbf{M}} &= \bordermatrix{%
 & C & D \cr
 C &\ \overline{b}-\overline{c},\ \overline{b}-\overline{c} & \ -\overline{c},\ \overline{b} \cr
 D &\ \overline{b},\ -\overline{c} & \ 0,\ 0 \cr
} ,
\end{align}
\end{linenomath}
where $\overline{b}=\frac{1}{N}\sum_{i=1}^{N}b_{i}$ and $\overline{c}=\frac{1}{N}\sum_{i=1}^{N}c_{i}$. Therefore, the dynamics of the asymmetric game are the same as those of its symmetric counterpart with benefit, $\overline{b}$, and cost, $\overline{c}$, regardless of network configuration or resource distribution. Under death-birth (resp. imitation) updating, this result implies that cooperation is expected to increase if and only if $\overline{b}/\overline{c}>k$ (resp. $\overline{b}/\overline{c}>k+2$), where $k$ is the degree of the (regular) network \citep{ohtsuki:Nature:2006}. \fig{Ecological}(A) compares the predicted result obtained from $\overline{\mathbf{M}}$ to simulation data for imitation updating when benefit and cost values are distributed according to Gaussian random variables.
\end{example}

\begin{figure}
\begin{center}
\includegraphics[width=0.5\textwidth]{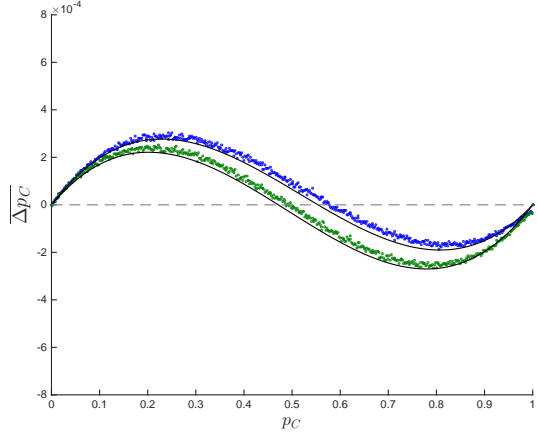}
\end{center}
\caption{Average change in the frequency of cooperators, $\overline{\Delta p_{C}}$, as a function of the frequency of cooperators, $p_{C}$, for a spatially non-additive Snowdrift Game, \eq{adaptedsnowdrift}, with selection intensity $\beta =0.01$. The blue and green data are obtained using pairwise comparison updating and differ only in the configuration of the underlying network, which in both cases is a random regular graph of size $N=500$ and degree $k=3$. Every vertex has a benefit value of $b=4.0$, and the cost values are split equally, with half of the vertices having $c_{1}=0.5$ and the remaining half having $c_{2}=5.5$. The average payoff for mutual cooperation, \eq{cnetwork}, is $3.069$ (blue) and $2.961$ (green), which suggests that the former arrangement is more attractive for cooperation. The analytical predictions (solid lines) are obtained from Eq. (\ref{expectationPC}) in Methods (and are linear in $\beta$ for $\beta\ll 1$).
\label{fig:NetworkDependence}}
\end{figure}

\begin{figure*}
\begin{center}
\includegraphics[width=1.0\textwidth]{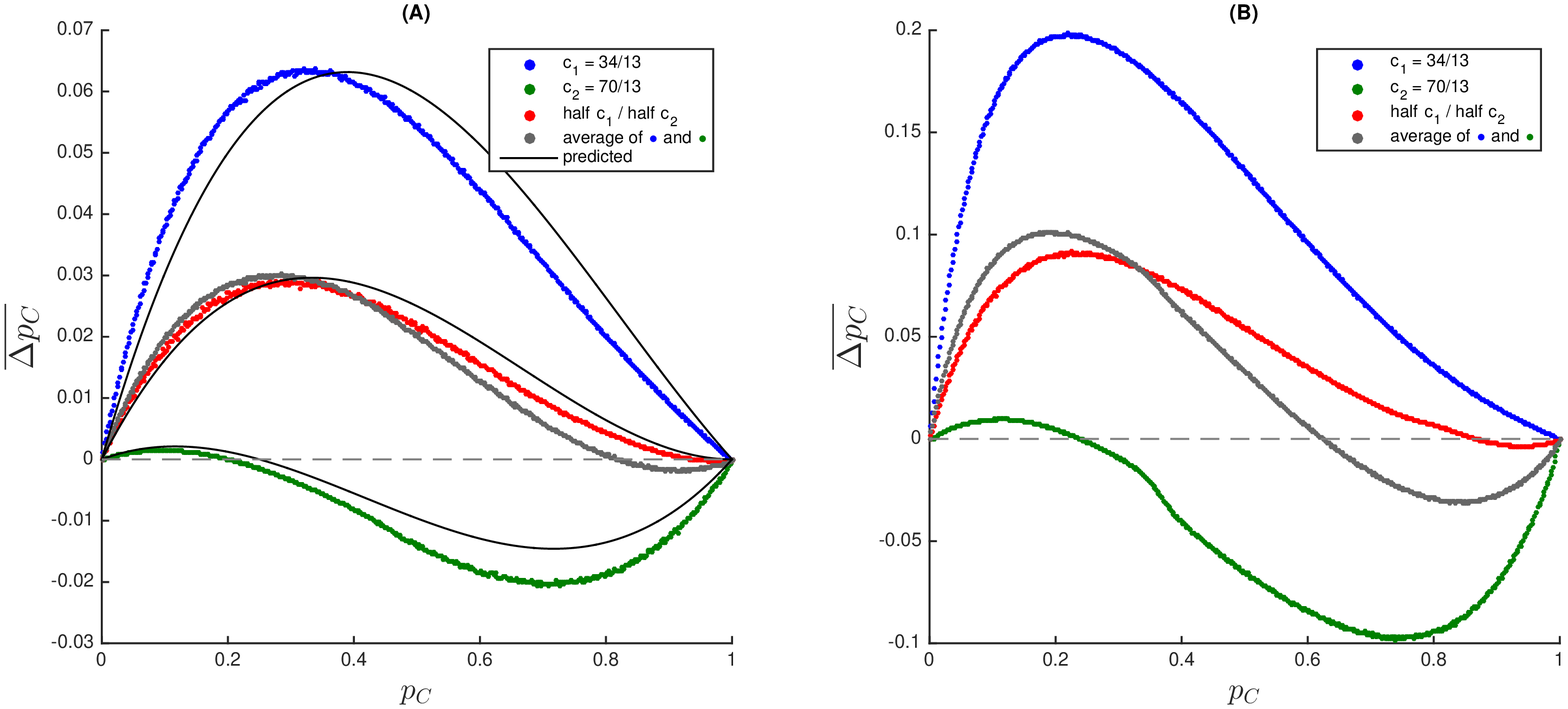}
\end{center}
\caption{The Snowdrift Games of Fig. \ref{fig:Ecological}(B) with the stronger selection strengths $\beta =0.1$ (A) and $\beta =0.5$ (B). For each of the three games (with benefit $b=5.0$ and costs $c_{1}$, $c_{2}$, and half $c_{1}$/half $c_{2}$, respectively), the simulation results differ from the prediction of pair approximation already for $\beta =0.1$ (A). Moreover, for $\beta =0.5$, (B) makes it clear that \thm{maintheorem} no longer holds since the average change in cooperators in the game with mixed costs (red) differs from the average (grey) of these changes for the games with costs $c_{1}$ only (blue) and $c_{2}$ only (green). Thus, \thm{maintheorem} is peculiar to weak selection.\label{fig:EcologicalStrongerSelection}}
\end{figure*}

\begin{example}[Snowdrift Game with ecological asymmetry]
In order to illustrate when \cor{spatiallyadditive} fails, we turn to cooperation in the Snowdrift Game \citep{hauert:Nature:2004,doebeli:EL:2005}. In this game, two drivers find themselves on either side of a snowdrift. If both cooperate in clearing the snowdrift, they share the cost, $c$, equally, and both receive the benefit of being able to pass, $b$. If one player cooperates and the other defects, both players receive $b$ but the cooperator pays the full cost, $c$. If both players defect, each receives no benefit and pays no cost. In order to incorporate ecological asymmetry, we assume that the benefits are all the same since they are derived from being able to pass in the absence of a snowdrift. On the other hand, the cost a player pays to clear the snowdrift may depend on his or her location: the snowdrift may appear on an incline, for example, in which case one player shovels with the gradient and the other player against it. Moreover, when two cooperators meet, they might clear unequal shares of the snowdrift. Thus, the payoff matrix for a player at location $i$ against a player at location $j$ should be of the form
\begin{linenomath}
\begin{align}\label{eq:adaptedsnowdrift}
\mathbf{M}^{ij}\left(\alpha_{ij}\right) &:= \bordermatrix{%
 & C & D \cr
 C &\ b-\alpha_{ij}c_{i},\ b-\alpha_{ji}c_{j} & \ b-c_{i},\ b \cr
 D &\ b,\ b-c_{j} & \ 0,\ 0 \cr
} ,
\end{align}
\end{linenomath}
where $0\leqslant\alpha_{ij}\leqslant 1$ and $\alpha_{ij}+\alpha_{ji}=1$ \citep{du:EL:2009}. Intuitively, when two cooperators face one other, they each begin to clear the snowdrift and stop once they meet; the quantity $\alpha_{ij}$ indicates the fraction of the snowdrift a cooperator at location $i$ clears before meeting the cooperator at location $j$. A natural choice for $\alpha_{ij}$ is
\begin{linenomath}
\begin{align}
\alpha_{ij} &= \frac{c_{j}}{c_{i}+c_{j}} ,
\end{align}
\end{linenomath}
which is the unique value that gives $\alpha_{ij}c_{i}=\alpha_{ji}c_{j}$ for each $i$ and $j$, ensuring that the game is \textit{fair}, i.e. that the cooperator with the higher cost clears a smaller portion of the snowdrift than the one with the lower cost. Averaging the payoff to one cooperator against another over all possible locations gives
\begin{linenomath}
\begin{align}\label{eq:cnetwork}
\frac{1}{kN}\sum_{i,j=1}^{N}w_{ij}\left( b-\alpha_{ij}c_{i} \right) &= b-\frac{1}{kN}\sum_{i,j=1}^{N}w_{ij}\left(\frac{c_{i}c_{j}}{c_{i}+c_{j}}\right) ,
\end{align}
\end{linenomath}
which is the upper-left entry of $\overline{\mathbf{M}}$. In contrast, the remaining three entries of $\overline{\mathbf{M}}$ do not depend on $\left(w_{ij}\right)_{1\leqslant i,j\leqslant N}$. Therefore, provided there are at least two locations with distinct cost values, the dynamics of an evolutionary process depend on the particular network configuration (\thm{maintheorem}). This network dependence is illustrated in \fig{NetworkDependence}.

Suppose now that we set $\alpha_{ij}\equiv 1/2$ to model ecological asymmetry in the Snowdrift Game; that is, if two cooperators meet, they each clear exactly half of the snowdrift. If there are two cost values in the population, $c_{1}$ and $c_{2}$, with $c_{1}<b<c_{2}<2b$, then a player who incurs a cost of $c_{1}$ finds it beneficial to cooperate against a defector, but a player who incurs a cost of $c_{2}$ would rather defect in this situation. Thus, based on the social dilemma implied by the ranking of the payoffs, a player who incurs a cost of $c_{1}$ for cooperating is always playing a Snowdrift Game while a player who incurs a cost of $c_{2}$ is always playing a Prisoner's Dilemma. It follows that ecological asymmetry can account for multiple social dilemmas being played within a single population, even if the players all use the same set of strategies ($C$ and $D$). The payoff matrices of this particular game are spatially additive, so, by \cor{spatiallyadditive}, the dynamics do not depend on the network configuration. If $q$ is the fraction of vertices with cost value $c_{1}$ then $\overline{c}=qc_{1}+\left(1-q\right) c_{2}$ is the average cost of cooperation for a particular location and the dynamics are the same as those of the symmetric Snowdrift Game in which the cost of clearing a snowdrift is $\overline{c}$ (see \fig{Ecological}(B)). Fig. \ref{fig:EcologicalStrongerSelection} demonstrates that this result does not extend to stronger selection strengths, so \thm{maintheorem} is unique to weak selection.

Based on \thm{maintheorem} and the relative rank of payoffs, the social dilemma defined by the asymmetric game (\ref{eq:adaptedsnowdrift}) (for general $\alpha_{ij}$) is a Prisoner's Dilemma if $b<\overline{c}$ and a Snowdrift Game if $b>\overline{c}$ when selection is weak. That is, microscopically, there is a mixture of Prisoner's Dilemmas and Snowdrift Games, but, macroscopically, the process behaves like just one of these social dilemmas. Consequently, although the dynamics of this evolutionary process may depend on the network configuration, the type of social dilemma implied by this game does not.
\end{example}

\subsection{Genotypic asymmetry}

Another form of asymmetry is based on the genotypes of the players rather than their locations. Each player in the population has one of $\ell$ possible genotypes, and these genotypes are enumerated by the set $\left\{1,\dots ,\ell\right\}$. For an $n$-strategy game, the payoff matrix for a player whose genotype is $u$ against a player whose genotype is $v$ is
\begin{linenomath}
\begin{align}\label{genotypicmatrix}
\mathbf{M}^{uv} &:= \bordermatrix{%
 & A_{1} & A_{2} & \cdots & A_{n} \cr
A_{1} &\ a_{11}^{uv}, a_{11}^{vu} & \ a_{12}^{uv}, a_{21}^{vu} & \ \cdots & \ a_{1n}^{uv}, a_{n1}^{vu} \cr
A_{2} &\ a_{21}^{uv}, a_{12}^{vu} & \ a_{22}^{uv}, a_{22}^{vu} & \ \cdots & \ a_{2n}^{uv}, a_{n2}^{vu} \cr
\ \vdots &\ \vdots & \ \vdots & \ \ddots & \ \vdots \cr
A_{n} &\ a_{n1}^{uv}, a_{1n}^{vu} & \ a_{n2}^{uv}, a_{2n}^{vu} & \ \cdots & \ a_{nn}^{uv}, a_{nn}^{vu} \cr
} .
\end{align}
\end{linenomath}

We explore genotypic asymmetry for cultural and genetic processes separately:

\subsubsection{Cultural updating}

If genotypic asymmetry is incorporated into a cultural process, then the genotypes of the players never change; only the strategies of the players are updated. In a structured population, it follows that each player's genotype may be associated with his or her location, and this association is an invariant of the process. Thus, if $u\left(i\right)$ denotes the genotype of the player at location $i$, then we may apply \thm{maintheorem} to the matrices defined by $\mathbf{M}^{ij}=\mathbf{M}^{u\left(i\right) u\left(j\right)}$ for each $i$ and $j$. In this sense, genotypic asymmetry may be ``reduced" to ecological asymmetry in evolutionary games with cultural update rules. Note that, unlike ecological asymmetry, genotypic asymmetry does not \textit{require} a structured population. However, one can always think of a population as structured (even in the well-mixed case), and doing so allows one to make sense of the ``locations" of the players and to apply \thm{maintheorem} to cultural processes with genotypic asymmetry.

\begin{example}[Donation Game with genotypic asymmetry and cultural updating]\label{ex:culturalDonation}
In the Donation Game, a cooperator of genotype $u$ donates $b_{u}$ at a cost of $c_{u}$. Defectors contribute no benefit and pay no cost, irrespective of genotype. Consider imitation updating on a large, regular network of degree $k$, and let $u\left(i\right)$ denote the genotype of the player at location $i$ (henceforth ``player $i$"). Suppose that player $i$ is a cooperator, player $j$ is a defector, and that player $i$ imitates player $j$ and becomes a cooperator. Despite this strategy change, the genotype of player $i$ is still $u\left(i\right)$, and the payoff matrix for player $i$ against player $j$ is still $\mathbf{M}^{u\left(i\right) u\left(j\right)}$. On the other hand, consider the same process but with the genotypic asymmetry replaced by ecological asymmetry (and with $\mathbf{M}^{ij}:=\mathbf{M}^{u\left(i\right) u\left(j\right)}$ as the payoff matrix for the player at location $i$ against the player at location $j$). Since the genotype of a player at a given location never changes in an imitation process, the process with ecological asymmetry is well-defined; that is, $\mathbf{M}^{ij}$ is independent of the dynamics of the process for each $i$ and $j$. Therefore, we may instead study the evolution of cooperation in the process with ecological asymmetry, and we already know from Example \ref{ex:donationEcological} that, in the limit of weak selection, the frequency of cooperators in this Donation Game is expected to increase if and only if $\left(k+2\right)\sum_{i=1}^{N}c_{u\left(i\right)}<\sum_{i=1}^{N}b_{u\left(i\right)}$.
\end{example}

In contrast, for genetic update rules, the asymmetry present due to differing genotypes can be removed completely if the genotypes of offspring are determined by genetic inheritance:

\subsubsection{Genetic updating}

Genetic update rules are defined by the ability of players to propagate their offspring to other locations in the population by means of births and deaths. In other words, there is a reproductive step in which genetic information is passed from parent(s) to child. Both the death-birth and birth-death processes have genetic update rules, but reproduction need not be clonal for the update rule to be genetic. If the genotypes of offspring are determined by genetic inheritance, then the strategy \textit{and} genotype at each location are updated simultaneously: if the offspring of a player whose genotype is $u$ and whose strategy is $A_{r}$ replaces a player whose genotype is $v$ and whose strategy is $A_{s}$, then $v$ is updated to $u$ and $A_{s}$ is updated to $A_{r}$ synchronously. Therefore, rather than treating genotypes and strategies separately, we may consider them together in the form of pairs, $\left(u, A_{r}\right)$, linking genotype and strategy. These pairs may be thought of as composite strategies of a larger evolutionary game whose payoff matrix, $\widetilde{\mathbf{M}}$, is defined by
\begin{linenomath}
\begin{align}
\widetilde{\mathbf{M}}_{\left(u, A_{r}\right) ,\left(v, A_{s}\right)} &:= a_{rs}^{uv}
\end{align}
\end{linenomath}
for genotypes, $u$ and $v$, and strategies, $A_{r}$ and $A_{s}$. The map
\begin{linenomath}
\begin{align}\label{strategyBlowup}
\Big\{ \mathbf{M}^{uv} \Big\}_{u,v=1}^{\ell} \longrightarrow \widetilde{\mathbf{M}}
\end{align}
\end{linenomath}
resolves a collection of $n\times n$ asymmetric payoff matrices with a single symmetric payoff matrix, $\widetilde{\mathbf{M}}$, of size $\ell n\times\ell n$. This argument holds for any population structure, so evolutionary processes with genotypic asymmetry that are based on genetic update rules can be studied in any setting in which there is a theory of symmetric games. For example, we may use the results from pair approximation on large, regular networks to study the Donation Game with genotypic asymmetry and genetic updating:

\begin{example}[Donation Game with genotypic asymmetry and genetic updating]
As in Example \ref{ex:culturalDonation}, a cooperator of genotype $u$ in the Donation Game donates $b_{u}$ at a cost of $c_{u}$. Defectors contribute no benefit and pay no cost, irrespective of genotype. For the death-birth and birth-death update rules, defectors may be modeled as cooperators whose benefit and costs are both $0$. In the larger symmetric game defined by (\ref{strategyBlowup}), it follows that there are $\ell +1$ distinct composite strategies: $\left(1,C\right)$, $\left(2,C\right)$, $\dots$, $\left(\ell ,C\right)$, and $D:=\left(\ell +1,C\right)$. For death-birth updating on a large, regular network of degree $k$, cooperators of genotype $u\in\left\{1,\dots ,\ell\right\}$ are expected to increase if and only if
\begin{linenomath}
\begin{align}\label{eq:genotypicDB}
k\left(c_{u}-\sum_{v=1}^{\ell}c_{v}p_{v}\right) < b_{u}-\sum_{v=1}^{\ell}b_{v}p_{v} ,
\end{align}
\end{linenomath}
where, for each $v\in\left\{1,\dots ,\ell\right\}$, $p_{v}$ denotes the frequency of cooperators of genotype $v$ (i.e. the frequency of strategy $\left(v,C\right)$ in the larger symmetric game). The terms $\sum_{v=1}^{\ell}b_{v}p_{v}$ and $\sum_{v=1}^{\ell}c_{v}p_{v}$ are the average population benefit and cost values, respectively. Therefore, the condition for the expected increase in cooperators of a particular genotype depends on the average level of cooperation within the population. Eq. (\ref{eq:genotypicDB}) may be thought of as an analogue of the `$b/c>k$' rule of \citet{ohtsuki:Nature:2006} with $b$ replaced by the ``benefit premium," $b_{u}-\sum_{v=1}^{\ell}b_{v}p_{v}$, and $c$ replaced by the ``cost premium," $c_{u}-\sum_{v=1}^{\ell}c_{v}p_{v}$.

In the birth-death process, on the other hand, cooperators of genotype $u\in\left\{1,\dots ,\ell\right\}$ are expected to increase if and only if
\begin{linenomath}
\begin{align}\label{eq:genotypicBD}
c_{u} < \sum_{v=1}^{\ell}c_{v}p_{v} .
\end{align}
\end{linenomath}
Interestingly, this condition is independent of the benefit values and says that cooperators of genotype $u\in\left\{1,\dots ,\ell\right\}$ increase in abundance if they incur, on average, smaller costs for cooperating than the other cooperators.

Eqs. (\ref{eq:genotypicDB}) and (\ref{eq:genotypicBD}) are obtained by noticing that the expected change in the frequency of cooperators of genotype $u$, $\mathbb{E}\left[\Delta p_{u}\right]$, is a positive multiple of $b_{u}-\sum_{v=1}^{\ell}b_{v}p_{v}-k\left(c_{u}-\sum_{v=1}^{\ell}c_{v}p_{v}\right)$ in the death-birth process and of $\sum_{v=1}^{\ell}c_{v}p_{v}-c_{u}$ in the birth-death process (see Eqs. (\ref{expectationDB}) and (\ref{expectationBD}) in Methods). In the birth-death process, it follows that the expected change in the frequency of cooperators of genotype $u$ is close to $0$ if $p_{u}$ is close to $1$, hence increases in cooperators who pay nonzero costs are necessarily transient.
\end{example}

\section{Discussion}

Asymmetric games naturally separate standard evolutionary update rules into cultural and genetic classes. This distinction is important because it captures biological differences that are not always apparent in models of evolution based on symmetric games. For example, consider a model player whose offspring \textit{replaces} a focal player and a model player whose strategy is \textit{imitated} by a focal player. For symmetric games, processes based on these two types of updates are mathematically identical; if asymmetry is present, then the fact that one update is genetic (replacement) and the other is cultural (imitation) becomes important. Thus, asymmetric games can highlight fundamental differences in evolutionary processes that are based on distinct update rules but happen to behave similarly when the underlying game is symmetric.

In order to incorporate into evolutionary games the asymmetries commonly studied in classical game theory, our focus has been on games with asymmetric \textit{payoffs}. Games with asymmetric payoffs arise naturally from different forms of interaction heterogeneity. Dependence of payoffs on the environment is a reasonable assumption when considering ecological variation \citep{maciejewski:JTB:2014}. Certain patches may provide resources or have drawbacks that influence a player's success when using a particular strategy \citep{kun:NC:2013}. Asymmetric interactions may also be the result of heterogeneity in the sizes or strengths of players \citep{maynardsmith:AB:1976,hauser:JTB:2014}. Whether the source of asymmetry is the environment or the players themselves, our model effectively resolves a collection of microscopically asymmetric interactions with a macroscopically symmetric game in the limit of weak selection. Figs. \ref{fig:Ecological} and \ref{fig:NetworkDependence} illustrate this result for three common update rules.

Similar forms of asymmetry have been studied previously in evolutionary game theory: \citet{szolnoki:EL:2007} consider asymmetry appearing in the update rule that results in ``attractive" and ``repulsive" players in the pairwise comparison process. For games with population structures defined by \textit{two} graphs (``interaction" and ``dispersal" graphs), \citet{ohtsuki:PRL:2007,ohtsuki:JTB:2007} show that the evolution of cooperation can be inhibited by asymmetry arising from differences in these two graphs. On the other hand, \citet{pacheco:PLoSCB:2009} show that heterogeneous population structures can promote the evolution of cooperation by effectively transforming a collection of microscopic social dilemmas into a global coordination game. This result is reminiscent of our \thm{maintheorem}, which relates the microscopic interactions to the global behavior of a process. Such heterogeneous population structures can result in asymmetric interactions even if the underlying game is symmetric \citep{maciejewski:PLoSCB:2014}. These models, although somewhat different from ours, demonstrate that asymmetry (in its many forms) has a remarkable effect on evolutionary dynamics.

Although genotypic asymmetry can always be reduced to a (larger) symmetric game under genetic update rules, this symmetric game can be of independent interest. For example, \eq{genotypicBD} shows that if cooperators vary in size or strength, then certain cooperators may increase in the Donation Game even under birth-death updating. In contrast, cooperation never increases in the absence of cooperator variation \cite{ohtsuki:Nature:2006}. Though defectors still eventually outcompete cooperators, the transient increase in cooperators suggests that other evolutionary processes with this form of asymmetry can behave in novel ways.

If both ecological and genotypic asymmetries are present, they can be handled separately: genotypic asymmetry is reduced to either (i) ecological asymmetry (if the update rule is cultural) or (ii) a symmetric game with more strategies (if the update rule is genetic). In either case, an evolutionary game with both ecological and genotypic asymmetries can be reduced to a game with ecological asymmetry only and hence \thm{maintheorem} applies. Our framework handles asymmetry resulting from varying baseline traits due to \textit{both} environment and genotype, which could be referred to as \textit{phenotypic} asymmetry.

The presence of ecological or genotypic asymmetry in an evolutionary process does not necessarily depend on the selection strength or update rule; these forms of asymmetry may be incorporated into many evolutionary processes. \thm{maintheorem}, which effectively reduces a game with ecological asymmetry to a particular symmetric game, is stated for four common update rules in evolutionary game theory. Fig. \ref{fig:EcologicalStrongerSelection} demonstrates (using the asymmetric Snowdrift Game) that this theorem is specific to \textit{weak} selection. That selection is weak is often a reasonable assumption when using evolutionary games to study populations of organisms with many traits. However, our study of the asymmetric Snowdrift Game for stronger selection strengths suggests that the behavior of asymmetric games is more complicated if selection is strong. Though more difficult to treat analytically, symmetric games under strong selection are worthy of further investigation.

Asymmetry is omnipresent in nature, and any framework that is used to model evolution should take into account possible sources of asymmetry. We have formally introduced ecological and genotypic asymmetries into evolutionary game theory and have studied these asymmetries in the limit of weak selection. Asymmetry has a natural place in the Donation Game and the Snowdrift Game, but our results are applicable to any general $n$-strategy matrix game. Our treatment of asymmetry highlights important differences between models of cultural and genetic evolution that are not apparent in the traditional setting of symmetric games. Ecological and genotypic asymmetries cover a wide variety of background variation observed in biological populations, and, as such, our framework enhances the modeling capacity of evolutionary games.

\section{Methods}

For the two genetic processes (death-birth and birth-death) and the two cultural processes (imitation and pairwise comparison) we consider, we treat ecologically asymmetric games on a large, regular network using pair approximation \citep{matsuda:PTP:1992,ohtsuki:Nature:2006}. We assume here that the degree of the network, $k$, is at least $3$. For $k=2$, the network is just a cycle, and we do not treat this case here. The detailed steps of each calculation are omitted but we include the main setups to allow for reconstruction of the reported results. We begin by recalling the way in which these four processes are defined (see eg. \citet{ohtsuki:JTB:2006}):
\begin{enumerate}

\item[(DB)] In the death-birth process, a player is selected uniformly at random from the population for death. A neighbor of the focal individual is then selected to reproduce with probability proportional to relative fitness, and the resulting offspring replaces the deceased player;

\item[(BD)] In the birth-death process, an individual is selected from the population for reproduction with probability proportional to relative fitness, and the offspring replaces a neighbor at random;

\item[(IM)] In the imitation process, an individual is chosen uniformly at random to evaluate his or her strategy. This focal individual either adopts a strategy of a neighbor (with probability proportional to that neighbor's relative fitness) or retains his or her original strategy (with probability proportional to own relative fitness);

\item[(PC)] In the pairwise comparison process, a focal individual is selected uniformly at random from the population to evaluate his or her strategy. A model individual is then chosen uniformly at random from the neighbors of the focal individual as a basis for comparison, and the focal player adopts the strategy of the model player with probability proportional to the model player's relative fitness.

\end{enumerate}

\subsection{Notation and general remarks}

Let $\mathcal{S}=\left\{A_{1},\dots ,A_{n}\right\}$ be the set of pure strategies available to each player and suppose that there are $N$ players on a regular network of size $N$ (i.e. every node is occupied). A strategy pair $\left(A_{r},A_{s}\right)$ means a choice of a player using strategy $A_{r}$ who has as a neighbor a player using strategy $A_{s}$. Let
\begin{linenomath}
\begin{subequations}
\begin{align}
p_{r} &:= \textrm{ frequency of players using strategy $A_{r}$} ; \\
p_{rs} &:= \textrm{ frequency of strategy pairs $\left(A_{r},A_{s}\right)$} ; \\
q_{s\vert r} &:= \textrm{ conditional probability of finding an $s$ player next to an $r$ player} .
\end{align}
\end{subequations}
\end{linenomath}
We will make repeated use of the following properties of these quantities:
\begin{linenomath}
\begin{subequations}
\begin{align}
&\sum_{r=1}^{n}p_{r} = \sum_{s=1}^{n}q_{s\vert r} = 1 ; \\
&p_{s}q_{r\vert s}=p_{rs}=p_{sr}=p_{r}q_{s\vert r} . \label{pairs}
\end{align}
\end{subequations}
\end{linenomath}
Strictly speaking, the equalities $p_{s}q_{r\vert s}=p_{rs}=p_{sr}=p_{r}q_{s\vert r}$ need not hold in general. As a pathological example, one may consider the network with two nodes and a single undirected link between these nodes. If the player on the first node uses $A_{r}$, the player on the second node uses $A_{s}$, and $r\neq s$, then $p_{rs}=1$ but $p_{s}=1/2$, which gives $q_{r\vert s}=2$. However, for large random regular graphs \citep{bollobas:CUP:2001}, condition (\ref{pairs}) holds approximately, and we will take this equality as given in what follows.

For $\mathcal{X}\in\left\{p_{r},p_{rs},q_{s\vert r}\right\}_{1\leqslant r,s\leqslant n}$, let $\mathbb{E}\left[\Delta\mathcal{X}\right]$ denote the expected change in $\mathcal{X}$ in one step of the process. A pair $\left(A_{r},i\right)$ denotes a player on vertex $i$ using strategy $A_{r}$. Given pairs $\left(A_{r},i\right)$ and $\left(A_{s},j\right)$, we denote by $\pi_{\left(A_{s},j\right)}\left(A_{r},i\right)$ the expected payoff to a player at vertex $j$ playing strategy $A_{s}$ given that they have as a neighbor an individual playing strategy $A_{r}$ at vertex $i$. If $\beta\geqslant 0$ is a parameter representing the intensity of selection, then payoff, $\pi$, is converted to fitness, $f_{\beta}\left(\pi\right)$, via
\begin{linenomath}
\begin{align}
f_{\beta}\left(\pi\right) &:= \exp\Big\{\beta\pi\Big\} .
\end{align}
\end{linenomath}
When defined in this way, fitness is always positive.

The main theorem we prove is the following:
\begin{repeatedtheorem}
In the limit of weak selection, the dynamics of the ecologically asymmetric death-birth, birth-death, imitation, and pairwise comparison processes on a large, regular network may be approximated by the dynamics of a \textit{symmetric} game with the same update rule and payoff matrix $\overline{\mathbf{M}}:=\frac{1}{kN}\sum_{i,j=1}^{N}w_{ij}\mathbf{M}^{ij}$, i.e.
\begin{linenomath}
\begin{align}
\overline{\mathbf{M}} &= \bordermatrix{%
 & A_{1} & A_{2} & \cdots & A_{n} \cr
A_{1} &\ \overline{a}_{11}, \overline{a}_{11} & \ \overline{a}_{12}, \overline{a}_{21} & \ \cdots & \ \overline{a}_{1n}, \overline{a}_{n1} \cr
A_{2} &\ \overline{a}_{21}, \overline{a}_{12} & \ \overline{a}_{22}, \overline{a}_{22} & \ \cdots & \ \overline{a}_{2n}, \overline{a}_{n2} \cr
\ \vdots &\ \vdots & \ \vdots & \ \ddots & \ \vdots \cr
A_{n} &\ \overline{a}_{n1}, \overline{a}_{1n} & \ \overline{a}_{n2}, \overline{a}_{2n} & \ \cdots & \ \overline{a}_{nn}, \overline{a}_{nn} \cr
} ,
\end{align}
\end{linenomath}
where $\overline{a}_{st}:=\frac{1}{kN}\sum_{i,j=1}^{N}w_{ij}a_{st}^{ij}$ for each $s$ and $t$.
\end{repeatedtheorem}

\thm{maintheorem} is established for each of these four update rules separately:
\subsection{Death-birth updating}

If an individual is playing strategy $A_{r}$ at node $i$, $A_{s}$ at $j$, and if $w_{ij}\neq 0$, then
\begin{linenomath}
\begin{align}
\pi_{\left(A_{s},j\right)}\left(A_{r},i\right) &= a_{sr}^{ji} + \sum_{m\neq i}w_{jm}\sum_{t=1}^{n}a_{st}^{jm}q_{t\vert s} .
\end{align}
\end{linenomath}
Suppose that an $\left(A_{r},i\right)$ individual is selected for death. The probability that $\left(A_{s},j\right)$ replaces this focal individual is proportional to $f_{\beta}\left(\pi_{\left(A_{s},j\right)}\left(A_{r},i\right)\right)$. For each $i$, let $\left(i_{1},\dots ,i_{k}\right)$ be an enumeration of the indices $j$ with $w_{ij}\neq 0$ (say, in increasing order) and let $s_{\ell}$ be the strategy used by the player at vertex $i_{\ell}$. If $\left(A_{r},i\right)$ is chosen for death, then the probability that it is replaced by $\left(A_{s_{\ell}},i_{\ell}\right)$ is
\begin{linenomath}
\begin{align}
\frac{f_{\beta}\left(\pi_{\left(A_{s_{\ell}},i_{\ell}\right)}\left(A_{r},i\right)\right)}{\sum_{j=1}^{k}f_{\beta}\left(\pi_{\left(A_{s_{i_{j}}},i_{j}\right)}\left(A_{r},i\right)\right)} .
\end{align}
\end{linenomath}
The Taylor expansion of this term for small $\beta$ is
\begin{linenomath}
\begin{align}
\frac{f_{\beta}\left(\pi_{\left(A_{s_{\ell}},i_{\ell}\right)}\left(A_{r},i\right)\right)}{\sum_{j=1}^{k}f_{\beta}\left(\pi_{\left(A_{s_{i_{j}}},i_{j}\right)}\left(A_{r},i\right)\right)} &= \frac{1}{k} + \beta\left(\frac{k\pi_{\left(A_{s},i_{\ell}\right)}\left(A_{r},i\right) -\sum_{j=1}^{k}\pi_{\left(A_{s_{i_{j}}},i_{j}\right)}\left(A_{r},i\right)}{k^{2}}\right) + O\left(\beta^{2}\right) .
\end{align}
\end{linenomath}
This expansion will be used frequently in the displays that follow.

\subsubsection{Approximation of the expected change in strategy frequencies}

Let $\delta_{x,y}$ be the Kronecker delta (defined to be $1$ if $x=y$ and $0$ otherwise). The probability of choosing the player on vertex $i$ for death is $1/N$. The chance that this player is using strategy $A_{h}$ is $p_{h}$. Suppose that $\left(A_{s_{i_{1}}},\dots ,A_{s_{i_{k}}}\right)$ is a $k$-tuple of strategies. If the focal player at vertex $i$ uses strategy $A_{h}$, then the probability that the player on vertex $i_{\ell}$ uses strategy $A_{s_{i_{\ell}}}$ for each $\ell =1,\dots ,k$ is $q_{s_{i_{1}}\vert h}\cdots q_{s_{i_{k}}\vert h}$. Thus,
\begin{linenomath}
\begin{align}
\mathbb{E}\left[\Delta p_{r}\right] &= \frac{1}{N}\sum_{i=1}^{N}\sum_{h\neq r}p_{h}\sum_{s_{i_{1}},\dots ,s_{i_{k}}=1}^{n}q_{s_{i_{1}}\vert h}\cdots q_{s_{i_{k}}\vert h}\sum_{\ell =1}^{k}\delta_{s_{i_{\ell}},r}\left(\frac{f_{\beta}\left(\pi_{\left(A_{r},i_{\ell}\right)}\left(A_{h},i\right)\right)}{\sum_{j=1}^{k}f_{\beta}\left(\pi_{\left(A_{s_{i_{j}}},i_{j}\right)}\left(A_{h},i\right)\right)}\right)\left(\frac{1}{N}\right) \nonumber \\
&\quad +\frac{1}{N}\sum_{i=1}^{N}p_{r}\sum_{s_{i_{1}},\dots ,s_{i_{k}}=1}^{n}q_{s_{i_{1}}\vert r}\cdots q_{s_{i_{k}}\vert r}\sum_{h\neq r}\sum_{\ell =1}^{k}\delta_{s_{i_{\ell}},h}\left(\frac{f_{\beta}\left(\pi_{\left(A_{h},i_{\ell}\right)}\left(A_{r},i\right)\right)}{\sum_{j=1}^{k}f_{\beta}\left(\pi_{\left(A_{s_{i_{j}}},i_{j}\right)}\left(A_{r},i\right)\right)}\right)\left(-\frac{1}{N}\right)
\end{align}
\end{linenomath}
for each strategy, $A_{r}$. The Taylor expansion to first-order yields
\begin{linenomath}
\begin{align}\label{dbstrategy}
\mathbb{E}\left[\Delta p_{r}\right] &\approx \beta\left(\frac{\left(k-1\right)p_{r}}{k^{2}N^{2}}\right)\Big(\left(\textrm{A}\right) - \left(\textrm{B}\right) -\left(\textrm{C}\right) + \left(\textrm{D}\right)\Big) + O\left(\beta^{2}\right) ,
\end{align}
\end{linenomath}
where
\begin{linenomath}
\begin{subequations}
\begin{align}
\left(\textrm{A}\right) &= \sum_{h\neq r}q_{h\vert r}\sum_{i=1}^{N}\sum_{\ell =1}^{k}\sum_{s_{i_{\ell}}=1}^{n}q_{s_{i_{\ell}}\vert r}\pi_{\left(A_{s_{i_{\ell}}},i_{\ell}\right)}\left(A_{r},i\right) ; \\
\left(\textrm{B}\right) &= \sum_{h\neq r}q_{h\vert r}\sum_{i=1}^{N}\sum_{\ell =1}^{k}\sum_{s_{i_{\ell}}=1}^{n}q_{s_{i_{\ell}}\vert h}\pi_{\left(A_{s_{i_{\ell}}},i_{\ell}\right)}\left(A_{h},i\right) ; \\
\left(\textrm{C}\right) &= \sum_{h\neq r}q_{h\vert r}\sum_{i=1}^{N}\sum_{\ell =1}^{k}\pi_{\left(A_{h},i_{\ell}\right)}\left(A_{r},i\right) ; \\
\left(\textrm{D}\right) &= \sum_{h\neq r}q_{h\vert r}\sum_{i=1}^{N}\sum_{\ell =1}^{k}\pi_{\left(A_{r},i_{\ell}\right)}\left(A_{h},i\right) .
\end{align}
\end{subequations}
\end{linenomath}

\subsubsection{Approximation of the expected change in pair frequencies}

If $r\neq s$, then
\begin{linenomath}
\begin{align}
\mathbb{E}\left[\Delta p_{rs}\right] &= \frac{1}{N}\sum_{i=1}^{N}\sum_{h\neq r,s}p_{h}\sum_{s_{i_{1}},\dots ,s_{i_{k}}=1}^{n}q_{s_{i_{1}}\vert h}\cdots q_{s_{i_{k}}\vert h} \nonumber \\
&\quad\quad \times\sum_{\ell =1}^{k}\delta_{s_{i_{\ell}},r}\left(\frac{f_{\beta}\left(\pi_{\left(A_{r},i_{\ell}\right)}\left(A_{h},i\right)\right)}{\sum_{j=1}^{k}f_{\beta}\left(\pi_{\left(A_{s_{i_{j}}},i_{j}\right)}\left(A_{h},i\right)\right)}\right)\left(\frac{2\sum_{\alpha =1}^{k}\delta_{s_{i_{\alpha}},s}}{kN}\right) \nonumber \\
&\quad + \frac{1}{N}\sum_{i=1}^{N}\sum_{h\neq r,s}p_{h}\sum_{s_{i_{1}},\dots ,s_{i_{k}}=1}^{n}q_{s_{i_{1}}\vert h}\cdots q_{s_{i_{k}}\vert h} \nonumber \\
&\quad\quad \times\sum_{\ell =1}^{k}\delta_{s_{i_{\ell}},s}\left(\frac{f_{\beta}\left(\pi_{\left(A_{s},i_{\ell}\right)}\left(A_{h},i\right)\right)}{\sum_{j=1}^{k}f_{\beta}\left(\pi_{\left(A_{s_{i_{j}}},i_{j}\right)}\left(A_{h},i\right)\right)}\right)\left(\frac{2\sum_{\alpha =1}^{k}\delta_{s_{i_{\alpha}},r}}{kN}\right) \nonumber \\
&\quad +\frac{1}{N}\sum_{i=1}^{N}p_{r}\sum_{s_{i_{1}},\dots ,s_{i_{k}}=1}^{n}q_{s_{i_{1}}\vert r}\cdots q_{s_{i_{k}}\vert r} \nonumber \\
&\quad\quad \times\sum_{\ell =1}^{k}\delta_{s_{i_{\ell}},s}\left(\frac{f_{\beta}\left(\pi_{\left(A_{s},i_{\ell}\right)}\left(A_{r},i\right)\right)}{\sum_{j=1}^{k}f_{\beta}\left(\pi_{\left(A_{s_{i_{j}}},i_{j}\right)}\left(A_{r},i\right)\right)}\right)\left(\frac{2\sum_{\alpha =1}^{k}\left(\delta_{s_{i_{\alpha}},r}-\delta_{s_{i_{\alpha}},s}\right)}{kN}\right) \nonumber \\
&\quad +\frac{1}{N}\sum_{i=1}^{N}p_{r}\sum_{s_{i_{1}},\dots ,s_{i_{k}}=1}^{n}q_{s_{i_{1}}\vert r}\cdots q_{s_{i_{k}}\vert r} \nonumber \\
&\quad\quad \times\sum_{h\neq r,s}\sum_{\ell =1}^{k}\delta_{s_{i_{\ell}},h}\left(\frac{f_{\beta}\left(\pi_{\left(A_{h},i_{\ell}\right)}\left(A_{r},i\right)\right)}{\sum_{j=1}^{k}f_{\beta}\left(\pi_{\left(A_{s_{i_{j}}},i_{j}\right)}\left(A_{r},i\right)\right)}\right)\left(-\frac{2\sum_{\alpha =1}^{k}\delta_{s_{i_{\alpha}},s}}{kN}\right) \nonumber \\
&\quad +\frac{1}{N}\sum_{i=1}^{N}p_{s}\sum_{s_{i_{1}},\dots ,s_{i_{k}}=1}^{n}q_{s_{i_{1}}\vert s}\cdots q_{s_{i_{k}}\vert s} \nonumber \\
&\quad\quad \times\sum_{\ell =1}^{k}\delta_{s_{i_{\ell}},r}\left(\frac{f_{\beta}\left(\pi_{\left(A_{r},i_{\ell}\right)}\left(A_{s},i\right)\right)}{\sum_{j=1}^{k}f_{\beta}\left(\pi_{\left(A_{s_{i_{j}}},i_{j}\right)}\left(A_{s},i\right)\right)}\right)\left(\frac{2\sum_{\alpha =1}^{k}\left(\delta_{s_{i_{\alpha}},s}-\delta_{s_{i_{\alpha}},r}\right)}{kN}\right) \nonumber \\
&\quad +\frac{1}{N}\sum_{i=1}^{N}p_{s}\sum_{s_{i_{1}},\dots ,s_{i_{k}}=1}^{n}q_{s_{i_{1}}\vert s}\cdots q_{s_{i_{k}}\vert s} \nonumber \\
&\quad\quad \times\sum_{h\neq r,s}\sum_{\ell =1}^{k}\delta_{s_{i_{\ell}},h}\left(\frac{f_{\beta}\left(\pi_{\left(A_{h},i_{\ell}\right)}\left(A_{s},i\right)\right)}{\sum_{j=1}^{k}f_{\beta}\left(\pi_{\left(A_{s_{i_{j}}},i_{j}\right)}\left(A_{s},i\right)\right)}\right)\left(-\frac{2\sum_{\alpha =1}^{k}\delta_{s_{i_{\alpha}},r}}{kN}\right) .
\end{align}
\end{linenomath}
On the other hand,
\begin{linenomath}
\begin{align}
\mathbb{E}\left[\Delta p_{rr}\right] &= \frac{1}{N}\sum_{i=1}^{n}\sum_{h\neq r}p_{h}\sum_{s_{i_{1}},\dots ,s_{i_{k}}=1}^{n}q_{s_{i_{1}}\vert h}\cdots q_{s_{i_{k}}\vert h} \nonumber \\
&\quad\quad \times\sum_{\ell =1}^{k}\delta_{s_{i_{\ell}},r}\left(\frac{f_{\beta}\left(\pi_{\left(A_{r},i_{\ell}\right)}\left(A_{h},i\right)\right)}{\sum_{j=1}^{k}f_{\beta}\left(\pi_{\left(A_{s_{i_{j}}},i_{j}\right)}\left(A_{h},i\right)\right)}\right)\left(\frac{2\sum_{\alpha =1}^{k}\delta_{s_{i_{\alpha}},r}}{kN}\right) \nonumber \\
&\quad +\frac{1}{N}\sum_{i=1}^{n}p_{r}\sum_{s_{i_{1}},\dots ,s_{i_{k}}=1}^{n}q_{s_{i_{1}}\vert r}\cdots q_{s_{i_{k}}\vert r} \nonumber \\
&\quad\quad \times\sum_{h\neq r}\sum_{\ell =1}^{k}\delta_{s_{i_{\ell}},h}\left(\frac{f_{\beta}\left(\pi_{\left(A_{h},i_{\ell}\right)}\left(A_{r},i\right)\right)}{\sum_{j=1}^{k}f_{\beta}\left(\pi_{\left(A_{s_{i_{j}}},i_{j}\right)}\left(A_{r},i\right)\right)}\right)\left(-\frac{2\sum_{\alpha =1}^{k}\delta_{s_{i_{\alpha}},r}}{kN}\right) .
\end{align}
\end{linenomath}
The zeroth-order Taylor expansion yields
\begin{linenomath}
\begin{align}\label{dbpairdifferent}
\mathbb{E}\left[\Delta p_{rs}\right] &\approx \frac{4p_{r}}{kN}\left(- kq_{s\vert r} + \left(k-1\right)\sum_{h=1}^{n}q_{s\vert h}q_{h\vert r}\right) +O\left(\beta\right)
\end{align}
\end{linenomath}
if $r\neq s$, and
\begin{linenomath}
\begin{align}\label{dbpairsame}
\mathbb{E}\left[\Delta p_{rr}\right] &\approx \frac{2p_{r}}{kN}\left(1-kq_{r\vert r} + \left(k-1\right)\sum_{h=1}^{k}q_{r\vert h}q_{h\vert r}\right) +O\left(\beta\right).
\end{align}
\end{linenomath}
Therefore, $\mathbb{E}\left[\Delta p_{r}\right] =O\left(\beta\right)$ (by Eq. (\ref{dbstrategy})) and $\mathbb{E}\left[\Delta p_{rs}\right] =O\left(1\right)$ (by Eqs. (\ref{dbpairdifferent}) and (\ref{dbpairsame})) for each $r$ and $s$, which results in a separation of timescales between the strategy frequencies and the pair frequencies. In particular, the pair frequencies will reach their equilibrium much more quickly than the strategy frequencies will, so we can examine the expression for $\mathbb{E}\left[\Delta p_{r}\right]$ under the assumption that the pair frequencies have reached their equilibrium \citep{ohtsuki:Nature:2006}.

\subsubsection{Weak-selection dynamics}\label{subsubsec:weakdynamics}

Assuming that each update takes place in one unit of time, we can approximate the dynamics by the deterministic systems $\dot{p}_{r}=\mathbb{E}\left[\Delta p_{r}\right]$ and $\dot{p}_{rs}=\mathbb{E}\left[\Delta p_{rs}\right]$ for each $r$ and $s$ \citep{ohtsuki:Nature:2006,ohtsuki:JTB:2006}. Since $\beta$ is small, we see that the latter system will reach equilibrium much quicker than the former. When the pair frequencies have reached equilibrium (i.e. $\mathbb{E}\left[\Delta p_{rs}\right] =0$), we have
\begin{linenomath}
\begin{align}
kq_{s\vert r} &= \delta_{s,r} + \left(k-1\right)\sum_{h=1}^{n}q_{s\vert h}q_{h\vert r} .
\end{align}
\end{linenomath}
\citet{ohtsuki:JTB:2006} show that this equation implies that
\begin{linenomath}
\begin{align}\label{localequilibrium}
q_{r\vert s} &= p_{r} + \left(\frac{1}{k-1}\right)\left(\delta_{s,r}-p_{r}\right) .
\end{align}
\end{linenomath}
Assuming the system has reached this local equilibrium, we then have
\begin{linenomath}
\begin{align}\label{expectationDB}
\mathbb{E}\left[\Delta p_{r}\right] &\approx \beta\left(\frac{\left(k-1\right)p_{r}}{k^{2}N^{2}}\right) \Bigg(\left(k+1\right)\sum_{i,j=1}^{N}w_{ij}\sum_{s=1}^{n}a_{rs}^{ij}q_{s\vert r} \nonumber \\
&\qquad\qquad\qquad\qquad\qquad -k\sum_{i,j=1}^{N}w_{ij}\sum_{s,t=1}^{n}a_{st}^{ij}q_{t\vert s}q_{s\vert r}-\sum_{i,j=1}^{N}w_{ij}\sum_{s,t=1}^{n}a_{st}^{ij}q_{s\vert t}q_{t\vert r}\Bigg) + O\left(\beta^{2}\right) \nonumber \\
&= \beta\left(\frac{\left(k-1\right)p_{r}}{kN}\right)\Bigg(\left(k+1\right)\sum_{s=1}^{n}\overline{a}_{rs}q_{s\vert r}-k\sum_{s,t=1}^{n}\overline{a}_{st}q_{t\vert s}q_{s\vert r}-\sum_{s,t=1}^{n}\overline{a}_{st}q_{s\vert t}q_{t\vert r}\Bigg) + O\left(\beta^{2}\right) \nonumber \\
&= \beta\left(\frac{\left(k-2\right) p_{r}}{k\left(k-1\right) N}\right) \Bigg(-\left(k-2\right)\left(k+1\right)\sum_{s,t=1}^{n}\overline{a}_{st}p_{s}p_{t}+\left(k^{2}-k-1\right)\sum_{s=1}^{n}\overline{a}_{rs}p_{s} \nonumber \\
&\qquad\qquad\qquad\qquad\qquad -\sum_{s=1}^{n}\overline{a}_{sr}p_{s}-\left(k+1\right)\sum_{s=1}^{n}\overline{a}_{ss}p_{s}+\left(k+1\right) \overline{a}_{rr}\Bigg) + O\left(\beta^{2}\right)
\end{align}
\end{linenomath}
as long as $\beta$ is small. Therefore, if we choose an appropriate time scale and set
\begin{linenomath}
\begin{subequations}
\begin{align}
\dot{p}_{r} &= \frac{\mathbb{E}\left[\Delta p_{r}\right]}{\Delta t} ; \\
\overline{b}_{rs} &= \frac{\overline{a}_{rr}+\overline{a}_{rs}-\overline{a}_{sr}-\overline{a}_{ss}}{k-2} ; \\
\phi &= \sum_{s,t=1}^{n}p_{s}p_{t}\overline{a}_{st} ,
\end{align}
\end{subequations}
\end{linenomath}
then $\dot{p}_{r}=p_{r}\left(\sum_{s=1}^{n}p_{s}\left(\overline{a}_{rs}+\overline{b}_{rs}\right) - \phi\right)$, recovering the replicator equation of \citet{ohtsuki:JTB:2006}. It follows that the dynamics depend on $\overline{\mathbf{M}}$, proving \thm{maintheorem} for death-birth updating.

\subsection{Birth-death updating}

In the birth-death process, an individual is selected for reproduction with probability proportional to relative fitness. The offspring of the selected player then replaces a random neighbor. Rather than trying to approximate the total fitness of the population, we will simply denote this value by $f^{\textrm{pop}}$. Since this value is positive, it does not influence the sign of the expectation values and as such we will largely ignore it. We have
\begin{linenomath}
\begin{align}
\mathbb{E}\left[\Delta p_{r}\right] &= \frac{1}{f^{\textrm{pop}}}Np_{r}\left(\frac{1}{N}\right)\sum_{i=1}^{N}\sum_{s_{i_{1}},\dots ,s_{i_{k}}=1}^{n}q_{s_{i_{1}}\vert r}\cdots q_{s_{i_{k}}\vert r}f_{\beta}\left(\sum_{\ell =1}^{k}a_{rs_{i_{\ell}}}^{ii_{\ell}}\right)\sum_{h\neq r}\left(\frac{\sum_{j=1}^{k}\delta_{s_{i_{j}},h}}{k}\right)\left(\frac{1}{N}\right) \nonumber \\
&\quad +\frac{1}{f^{\textrm{pop}}}\sum_{h\neq r}Np_{h}\left(\frac{1}{N}\right)\sum_{i=1}^{N}\sum_{s_{i_{1}},\dots ,s_{i_{k}}=1}^{n}q_{s_{i_{1}}\vert h}\cdots q_{s_{i_{k}}\vert h}f_{\beta}\left(\sum_{\ell =1}^{k}a_{hs_{i_{\ell}}}^{ii_{\ell}}\right)\left(\frac{\sum_{j=1}^{k}\delta_{s_{i_{j}},r}}{k}\right)\left(-\frac{1}{N}\right) .
\end{align}
\end{linenomath}

The local equilibrium conditions for birth-death updating turn out to be the same as those for death-birth updating (Eq. (\ref{localequilibrium})). These local equilibrium conditions do not take into account selection as long as $\beta$ is close to $0$, so they are essentially based on a neutral process in which at most one strategy is update at each time step. Therefore, it is perhaps not surprising that these conditions are the same for different processes based on one strategy update in each time step.

In the following expressions, by $x\propto y$ we mean that $x$ is proportional to $y$ with \textit{positive} constant of proportionality. Letting $\beta\rightarrow 0$ and using the local equilibrium conditions (as well as the same separation-of-timescales argument we used in \S\ref{subsubsec:weakdynamics}), we find that
\begin{linenomath}
\begin{align}\label{expectationBD}
\mathbb{E}\left[\Delta p_{r}\right] &\propto \beta p_{r}\left(k\sum_{i,j=1}^{N}w_{ij}\sum_{s=1}^{n}a_{rs}^{ij}q_{s\vert r}-\left(k-1\right)\sum_{i,j=1}^{N}w_{ij}\sum_{s,t=1}^{n}a_{st}^{ij}q_{t\vert s}q_{s\vert r}-\sum_{i,j=1}^{N}w_{ij}\sum_{s=1}^{n}a_{sr}^{ij}q_{s\vert r}\right) +O\left(\beta^{2}\right) \nonumber \\
&\propto \beta p_{r}\left(k\sum_{s=1}^{n}\overline{a}_{rs}q_{s\vert r}-\left(k-1\right)\sum_{s,t=1}^{n}\overline{a}_{st}q_{t\vert s}q_{s\vert r}-\sum_{s=1}^{n}\overline{a}_{sr}q_{s\vert r}\right) +O\left(\beta^{2}\right) \nonumber \\
&\propto \beta p_{r}\Bigg( -\left(k-2\right)\sum_{s,t=1}^{n}\overline{a}_{st}p_{s}p_{t} + \left(k-1\right)\sum_{s=1}^{n}\overline{a}_{rs}p_{s} - \sum_{s=1}^{n}\overline{a}_{sr}p_{s} - \sum_{s=1}^{n}\overline{a}_{ss}p_{s}+ \overline{a}_{rr} \Bigg) + O\left(\beta^{2}\right) .
\end{align}
\end{linenomath}
Just as we saw with the death-birth process, after choosing an appropriate time scale and letting
\begin{linenomath}
\begin{subequations}
\begin{align}
\overline{b}_{rs} &= \frac{\left(k+1\right)\overline{a}_{rr}+\overline{a}_{rs}-\overline{a}_{sr}-\left(k+1\right)\overline{a}_{ss}}{\left(k-2\right)\left(k+1\right)} ; \\
\phi &= \sum_{s,t=1}^{n}p_{s}p_{t}\overline{a}_{st} ,
\end{align}
\end{subequations}
\end{linenomath}
we have $\dot{p}_{r}=p_{r}\left(\sum_{s=1}^{n}p_{s}\left(\overline{a}_{rs}+\overline{b}_{rs}\right) - \phi\right)$, proving \thm{maintheorem} for birth-death updating.

\subsection{Imitation updating}

In the imitation process, an individual is selected uniformly at random from the population to evaluate his strategy. The chosen player then compares his fitness with the fitness of each neighbor and either adopts a new strategy or retains his or her current strategy (with probability proportional to relative fitness). Suppose that an individual at vertex $i$, playing $A_{r}$, is selected to evaluate his or her strategy. If $s\neq r$, then the probability that he or she adopts strategy $s$ is
\begin{linenomath}
\begin{align}
\frac{\sum_{\ell =1}^{k}\delta_{s_{\ell},s}f_{\beta}\left(\pi_{\left(A_{s_{\ell}},i_{\ell}\right)}\left(A_{r},i\right)\right)}{\sum_{j=1}^{k}f_{\beta}\left(\pi_{\left(A_{s_{i_{j}}},i_{j}\right)}\left(A_{r},i\right)\right) +f_{\beta}\left(\sum_{j=1}^{k}a_{rs_{i_{j}}}^{ii_{j}}\right)}
\end{align}
\end{linenomath}
and the probability that his strategy remains unchanged is
\begin{linenomath}
\begin{align}
\frac{\sum_{\ell =1}^{k}\delta_{s_{\ell},r}f_{\beta}\left(\pi_{\left(A_{s_{\ell}},i_{\ell}\right)}\left(A_{r},i\right)\right) +f_{\beta}\left(\sum_{j=1}^{k}a_{rs_{i_{j}}}^{ii_{j}}\right)}{\sum_{j=1}^{k}f_{\beta}\left(\pi_{\left(A_{s_{i_{j}}},i_{j}\right)}\left(A_{r},i\right)\right) +f_{\beta}\left(\sum_{j=1}^{k}a_{rs_{i_{j}}}^{ii_{j}}\right)} .
\end{align}
\end{linenomath}
We let $\pi_{\left(A_{s},j\right)}\left(A_{r},i\right)$ be the same as it was for death-birth updating. For small $\beta$,
\begin{linenomath}
\begin{align}
&\frac{f_{\beta}\left(\pi_{\left(A_{s_{\ell}},i_{\ell}\right)}\left(A_{r},i\right)\right)}{\sum_{j=1}^{k}f_{\beta}\left(\pi_{\left(A_{s_{i_{j}}},i_{j}\right)}\left(A_{r},i\right)\right) +f_{\beta}\left(\sum_{j=1}^{k}a_{rs_{i_{j}}}^{ii_{j}}\right)} \nonumber \\
&\quad\approx \frac{1}{k+1} + \beta\left(\frac{\left(k+1\right)\pi_{\left(A_{s_{\ell}},i_{\ell}\right)}\left(A_{r},i\right) -\sum_{j=1}^{k}\pi_{\left(A_{s_{i_{j}}},i_{j}\right)}\left(A_{r},i\right) -\sum_{j=1}^{k}a_{rs_{i_{j}}}^{ii_{j}}}{\left(k+1\right)^{2}}\right) +O\left(\beta^{2}\right) .
\end{align}
\end{linenomath}

\subsubsection{Approximation of the expected change in strategy frequencies}

For $r\in\left\{1,\dots ,n\right\}$,
\begin{linenomath}
\begin{align}
\mathbb{E}\left[\Delta p_{r}\right] &= \frac{1}{N}\sum_{i=1}^{N}\sum_{h\neq r}p_{h}\sum_{s_{i_{1}},\dots ,s_{i_{k}}=1}^{n}q_{s_{i_{1}}\vert h}\cdots q_{s_{i_{k}}\vert h} \nonumber \\
&\quad\quad \times\sum_{\ell =1}^{k}\delta_{s_{i_{\ell}},r}\left(\frac{f_{\beta}\left(\pi_{\left(A_{r},i_{\ell}\right)}\left(A_{h},i\right)\right)}{\sum_{j=1}^{k}f_{\beta}\left(\pi_{\left(A_{s_{i_{j}}},i_{j}\right)}\left(A_{h},i\right)\right) +f_{\beta}\left(\sum_{j=1}^{k}a_{hs_{i_{j}}}^{ii_{j}}\right)}\right)\left(\frac{1}{N}\right) \nonumber \\
&\quad +\frac{1}{N}\sum_{i=1}^{N}p_{r}\sum_{s_{i_{1}},\dots ,s_{i_{k}}=1}^{n}q_{s_{i_{1}}\vert r}\cdots q_{s_{i_{k}}\vert r} \nonumber \\
&\quad\quad \times\sum_{h\neq r}\sum_{\ell =1}^{k}\delta_{s_{i_{\ell}},h}\left(\frac{f_{\beta}\left(\pi_{\left(A_{h},i_{\ell}\right)}\left(A_{r},i\right)\right)}{\sum_{j=1}^{k}f_{\beta}\left(\pi_{\left(A_{s_{i_{j}}},i_{j}\right)}\left(A_{r},i\right)\right) +f_{\beta}\left(\sum_{j=1}^{k}a_{rs_{i_{j}}}^{ii_{j}}\right)}\right)\left(-\frac{1}{N}\right) .
\end{align}
\end{linenomath}
The local equilibrium conditions are exactly the same as they were for the death-birth process. Assuming that the system has reached this local equilibrium, the separation-of-timescales argument we used in \S\ref{subsubsec:weakdynamics} gives
\begin{linenomath}
\begin{align}\label{expectationIM}
\mathbb{E}\left[\Delta p_{r}\right] &\approx \beta\left(\frac{p_{r}}{\left(k+1\right)^{2}N^{2}}\right) \Bigg(\left(k^{2}+2k-1\right)\sum_{i,j=1}^{N}w_{ij}\sum_{s=1}^{n}a_{rs}^{ij}q_{s\vert r} \nonumber \\
&\qquad\qquad\qquad\qquad\qquad -\left(k^{2}+k-2\right)\sum_{i,j=1}^{N}w_{ij}\sum_{s,t=1}^{n}a_{st}^{ij}q_{t\vert s}q_{s\vert r} \nonumber \\
&\qquad\qquad\qquad\qquad\qquad -\left(k-1\right)\sum_{i,j=1}^{N}w_{ij}\sum_{s,t=1}^{n}a_{ts}^{ij}q_{t\vert s}q_{s\vert r} -2\sum_{i,j=1}^{N}w_{ij}\sum_{s=1}^{n}a_{sr}^{ij}q_{s\vert r}\Bigg) +O\left(\beta^{2}\right) \nonumber \\
&= \beta\left(\frac{kp_{r}}{\left(k+1\right)^{2}N}\right) \Bigg(\left(k^{2}+2k-1\right)\sum_{s=1}^{n}\overline{a}_{rs}q_{s\vert r} -\left(k^{2}+k-2\right)\sum_{s,t=1}^{n}\overline{a}_{st}q_{t\vert s}q_{s\vert r} \nonumber \\
&\qquad\qquad\qquad\qquad\qquad -\left(k-1\right)\sum_{s,t=1}^{n}\overline{a}_{ts}q_{t\vert s}q_{s\vert r} -2\sum_{s=1}^{n}\overline{a}_{sr}q_{s\vert r}\Bigg) +O\left(\beta^{2}\right) \nonumber \\
&= \beta\left(\frac{k\left(k-2\right) p_{r}}{\left(k-1\right)\left(k+1\right)^{2}N}\right) \Bigg( -\left(k-2\right)\left(k+3\right) \sum_{s,t=1}^{n}\overline{a}_{st}p_{s}p_{t}+\left(k^{2}+k-3\right)\sum_{s=1}^{n}\overline{a}_{rs}p_{s} \nonumber \\
&\qquad\qquad\qquad\qquad\qquad -3\sum_{s=1}^{n}\overline{a}_{sr}p_{s} -\left(k+3\right)\sum_{s=1}^{n}\overline{a}_{ss}p_{s} +\left(k+3\right)\overline{a}_{rr}\Bigg) + O\left(\beta^{2}\right) .
\end{align}
\end{linenomath}
With $\overline{b}_{rs}=\frac{\left(k+3\right)\overline{a}_{rr}+3\overline{a}_{rs}-3\overline{a}_{sr}-\left(k+3\right)\overline{a}_{ss}}{\left(k-2\right)\left(k+3\right)}$ and $\phi =\sum_{s,t=1}^{n}p_{s}p_{t}\overline{a}_{st}$, we have
\begin{linenomath}
\begin{align}
\dot{p}_{r} &= p_{r}\left(\sum_{s=1}^{n}p_{s}\left(\overline{a}_{rs}+\overline{b}_{rs}\right) - \phi\right) ,
\end{align}
\end{linenomath}
which establishes \thm{maintheorem} for imitation updating.

\subsection{Pairwise comparison updating}

In the pairwise comparison process, a focal individual is selected uniformly at random from the population. A model individual is then chosen uniformly at random from the neighbors of the focal individual. If $\pi_{\textrm{f}}$ and $\pi_{\textrm{m}}$ denote the payoffs to the focal and model individuals, respectively, then the focal player will adopt the strategy of the model player with probability
\begin{linenomath}
\begin{align}
\frac{1}{1+e^{\beta\left(\pi_{\textrm{f}}-\pi_{\textrm{m}}\right)}} &= \frac{f_{\beta}\left(\pi_{\textrm{m}}\right)}{f_{\beta}\left(\pi_{\textrm{m}}\right)+f_{\beta}\left(\pi_{\textrm{f}}\right)} ,
\end{align}
\end{linenomath}
where $\beta\geqslant 0$ is a real parameter representing the intensity of selection. In addition to the expected payoff $\pi_{\left(A_{s},j\right)}\left(A_{r},i\right)$ (defined in the same way as for death-birth updating), we let
\begin{linenomath}
\begin{align}
\pi_{\left(A_{s},i\right)} &:= \sum_{j=1}^{k}a_{ss_{i_{j}}}^{ii_{j}}
\end{align}
\end{linenomath}
if $\left(A_{s},i\right)$ has as a neighborhood $\left(A_{s_{i_{1}}},\dots ,A_{s_{i_{k}}}\right)$. With this notation in place, we have
\begin{linenomath}
\begin{align}
\mathbb{E}\left[\Delta p_{r}\right] &= \frac{1}{N}\sum_{i=1}^{N}\sum_{h\neq r}p_{h}\sum_{s_{i_{1}},\dots ,s_{i_{k}}=1}^{n}q_{s_{i_{1}}\vert h}\cdots q_{s_{i_{k}}\vert h} \nonumber \\
&\quad\quad \times\sum_{\ell =1}^{k}\left(\frac{1}{k}\right)\delta_{s_{i_{\ell}},r}\left(\frac{f_{\beta}\left(\pi_{\left(A_{r},i_{\ell}\right)}\left(A_{h},i\right)\right)}{f_{\beta}\left(\pi_{\left(A_{r},i_{\ell}\right)}\left(A_{h},i\right)\right)+f_{\beta}\left(\pi_{\left(A_{h},i\right)}\right)}\right)\left(\frac{1}{N}\right) \nonumber \\
&\quad +\frac{1}{N}\sum_{i=1}^{N}p_{r}\sum_{s_{i_{1}},\dots ,s_{i_{k}}=1}^{n}q_{s_{i_{1}}\vert r}\cdots q_{s_{i_{k}}\vert r} \nonumber \\
&\quad\quad \times\sum_{h\neq r}\sum_{\ell =1}^{k}\left(\frac{1}{k}\right)\delta_{s_{i_{\ell}},h}\left(\frac{f_{\beta}\left(\pi_{\left(A_{h},i_{\ell}\right)}\left(A_{r},i\right)\right)}{f_{\beta}\left(\pi_{\left(A_{h},i_{\ell}\right)}\left(A_{r},i\right)\right)+f_{\beta}\left(\pi_{\left(A_{r},i\right)}\right)}\right)\left(-\frac{1}{N}\right) .
\end{align}
\end{linenomath}
As $\beta\rightarrow 0$, we have
\begin{linenomath}
\begin{align}
\frac{f_{\beta}\left(x\right)}{f_{\beta}\left(x\right) +f_{\beta}\left(y\right)} &\approx \frac{1}{2} + \beta\left(\frac{x-y}{4}\right) + O\left(\beta^{2}\right) .
\end{align}
\end{linenomath}
Consequently, in the limit of weak selection,
\begin{linenomath}
\begin{align}\label{expectationPC}
\mathbb{E}\left[\Delta p_{r}\right] &\approx \beta\frac{p_{r}}{2kN^{2}}\left(k\sum_{i,j=1}^{N}w_{ij}\sum_{s=1}^{n}a_{rs}^{ij}q_{s\vert r} -\left(k-1\right)\sum_{i,j=1}^{N}w_{ij}\sum_{s,t=1}^{n}a_{st}^{ij}q_{t\vert s}q_{s\vert r} -\sum_{i,j=1}^{N}w_{ij}\sum_{s=1}^{n}a_{sr}^{ij}q_{s\vert r}\right) +O\left(\beta^{2}\right) \nonumber \\
&= \beta\frac{p_{r}}{2N}\left(k\sum_{s=1}^{n}\overline{a}_{rs}q_{s\vert r} -\left(k-1\right)\sum_{s,t=1}^{n}\overline{a}_{st}q_{t\vert s}q_{s\vert r} -\sum_{s=1}^{n}\overline{a}_{sr}q_{s\vert r}\right) +O\left(\beta^{2}\right) \nonumber \\
&= \beta\left(\frac{\left(k-2\right) p_{r}}{2\left(k-1\right) N}\right) \Bigg(-\left(k-2\right)\sum_{s,t=1}^{n}\overline{a}_{st}p_{s}p_{t}+\left(k-1\right)\sum_{s=1}^{n}\overline{a}_{rs}p_{s} \nonumber \\
&\qquad\qquad\qquad\qquad\qquad -\sum_{s=1}^{n}\overline{a}_{sr}p_{s}-\sum_{s=1}^{n}\overline{a}_{ss}p_{s}+\overline{a}_{rr}\Bigg) + O\left(\beta^{2}\right) .
\end{align}
\end{linenomath}
The local equilibrium conditions are exactly the same as they were for the other processes, but in this case they are not needed to arrive at this last expression for $\mathbb{E}\left[\Delta p_{r}\right]$. With $\overline{b}_{rs}=\frac{\overline{a}_{rr}+\overline{a}_{rs}-\overline{a}_{sr}-\overline{a}_{ss}}{k-2}$ and $\phi =\sum_{s,t=1}^{n}p_{s}p_{t}\overline{a}_{st}$, we have $\dot{p}_{r}=p_{r}\left(\sum_{s=1}^{n}p_{s}\left(\overline{a}_{rs}+\overline{b}_{rs}\right) - \phi\right)$. It follows that the dynamics of the pairwise comparison process depend on $\overline{\mathbf{M}}$, which completes the proof of \thm{maintheorem}.

Finally, we show that the dynamics of each process are independent of the particular network configuration if the asymmetric game is \textit{spatially additive}:
\begin{repeateddefinition}
If $a_{rs}^{ij}=x_{rs}^{i}+y_{rs}^{j}$ for each $r$ and $s$, then $\mathbf{M}^{ij}$ is called a \textit{spatially additive} payoff matrix. If $\mathbf{M}^{ij}$ is spatially additive for each $i$ and $j$, then the game is said to be spatially additive.
\end{repeateddefinition}

\begin{repeatedcorollary}\label{spatiallyadditive}
If $\mathbf{M}^{ij}$ is spatially additive for each $i$ and $j$, then the expected change in the frequency of strategy $A_{r}$, $\mathbb{E}\left[\Delta p_{r}\right]$, is independent of $\left(w_{ij}\right)_{1\leqslant i,j\leqslant N}$ for each $r$. In particular, the dynamics of the process do not depend on the particular network configuration.
\end{repeatedcorollary}
\begin{proof}
If $a_{rs}^{ij}=x_{rs}^{i}+y_{rs}^{j}$ for each $r,s,i,j$, then
\begin{linenomath}
\begin{align}
\overline{a}_{st} &= \frac{1}{kN}\sum_{i,j=1}^{N}w_{ij}a_{st}^{ij} = \frac{1}{N}\sum_{i=1}^{N}x_{rs}^{i} + \frac{1}{N}\sum_{j=1}^{N}y_{rs}^{j} ,
\end{align}
\end{linenomath}
which is independent of $\left(w_{ij}\right)_{1\leqslant i,j\leqslant N}$. The corollary then follows directly from \thm{maintheorem}.
\end{proof}

\subsection{Computer simulations}

In each simulation, a random $k$-regular network (with $k=3$) of $N=500$ vertices is generated. The selection intensity is $\beta =0.01$ for Figs. \ref{fig:Ecological} and \ref{fig:NetworkDependence}, $\beta =0.1$ for Fig. \ref{fig:EcologicalStrongerSelection}(A), and $\beta =0.5$ for Fig. \ref{fig:EcologicalStrongerSelection}(B). The figures are generated based on data collected from a number of cycles: In each cycle, the network is given an initial configuration of cooperators by first choosing a density, $d$, uniformly at random from the interval $\left[0,1\right]$, and then placing a cooperator (resp. defector) at each vertex with probability $d$ (resp. $1-d$). The update rule is applied until either $C$ or $D$ fixates. (The \textit{absorption time} depends on a number of factors including the game, selection strength, and initial configuration of the population.) Let $p_{C}\left(t\right)$ denote the frequency of cooperators at time $t$; $p_{C}\left(0\right)$ is just the initial frequency of cooperators. The frequency $p_{C}\left(t+1\right)$ is obtained from $p_{C}\left(t\right)$ by adding to it the change in the frequency of cooperators over the next $N$ ($=500$) updates. For each $t$, the quantity $p_{C}\left(t+1\right)-p_{C}\left(t\right)$ is associated with $p_{C}\left(t\right)$. Once $p_{C}\in\left\{0,1\right\}$, a new initial configuration of cooperators is chosen and the process is repeated. After each possible value of $p_{C}$ has at least $10^{5}$ associated data points (changes in cooperator frequency), these changes are averaged, and this resulting quantity, $\overline{\Delta p_{C}}$, is paired with the corresponding value of $p_{C}$. These pairs are then plotted to obtain Figs. \ref{fig:Ecological}, \ref{fig:NetworkDependence}, and \ref{fig:EcologicalStrongerSelection}. The results from pair approximation apply to the expected change over $\textit{one}$ update, but we can easily get a predicted result over $N$ updates (i.e. one Monte Carlo step) by scaling the expressions for $\mathbb{E}\left[\Delta p_{C}\right]$ by a factor of $N$.

Small deviations from the expected results are seen in each of the figures, and these deviations are due to the effects of finite selection parameter ($\beta$) and the finiteness of the set of possible values of $p_{C}$ ($\Delta p_{C}$ is a multiple of $1/N$). As an example of how these properties can give rise to small deviations, consider the Donation Game under imitation updating in Fig. \ref{fig:Ecological}(A). Eq. (\ref{expectationIM}) predicts that $\mathbb{E}\left[\Delta p_{C}\right]$ is always positive, yet we observe in Fig. \ref{fig:Ecological}(A) that this change becomes negative as $p_{C}\rightarrow 0,1$. If $p_{C}=\left(N-1\right) /N$ and $\beta >0$, then the only defector in the population has a higher payoff than all of the other cooperators. Let $f_{\beta}^{\left(j\right)}$ denote the fitness of the player at location $j$. Thus, with just a single defector (at location $i$) in a population of cooperators, we have $f_{\beta}^{\left(i\right)}\geqslant f_{\beta}^{\left(j\right)}$ for each $j\neq i$, with equality if and only if $\beta =0$. The expected change in the frequency of cooperators in the next time step is
\begin{linenomath}
\begin{align}\label{expchangeone}
\mathbb{E}\left[\Delta p_{C}\right] &= \left(\frac{1}{N}\right)\left(\frac{1}{N}\right)\left(1-\frac{f_{\beta}^{\left(i\right)}}{f_{\beta}^{\left(i\right)}+\sum_{\left\{j\ :\ w_{ij}=1\right\}}f_{\beta}^{\left(j\right)}}\right) \nonumber \\
&\quad -\left(\frac{1}{N}\right)\sum_{\left\{j\ :\ w_{ij}=1\right\}}\left(\frac{1}{N}\right)\left(\frac{f_{\beta}^{\left(i\right)}}{f_{\beta}^{\left(j\right)}+\sum_{\left\{l\ :\ w_{jl}=1\right\}}f_{\beta}^{\left(l\right)}}\right) .
\end{align}
\end{linenomath}
The first (resp. second) summation runs over all of the neighbors of $i$ (resp. $j$). For each $j\neq i$,
\begin{linenomath}
\begin{subequations}
\begin{align}
\frac{f_{\beta}^{\left(i\right)}}{f_{\beta}^{\left(i\right)}+\sum_{\left\{j\ :\ w_{ij}=1\right\}}f_{\beta}^{\left(j\right)}} &\geqslant \frac{1}{k+1} ; \\
\frac{f_{\beta}^{\left(i\right)}}{f_{\beta}^{\left(j\right)}+\sum_{\left\{l\ :\ w_{jl}=1\right\}}f_{\beta}^{\left(l\right)}} &\geqslant \frac{1}{k+1} ,
\end{align}
\end{subequations}
\end{linenomath}
both with equality if and only if $\beta =0$. Therefore, we see that
\begin{linenomath}
\begin{align}
\mathbb{E}\left[\Delta p_{C}\right] &\leqslant \left(\frac{1}{N}\right)\left(\frac{1}{N}\right)\left(1-\frac{1}{k+1}\right) - \left(\frac{1}{N}\right)\left(\frac{k}{N}\right)\left(\frac{1}{k+1}\right) = 0
\end{align}
\end{linenomath}
with equality if and only if $\beta =0$. The same argument explains the negative average changes as $p_{C}\rightarrow 0$. Since $p_{C}$ can only take on finitely many values for a given population size, similar arguments explain the small discrepancies between the actual and expected results for intermediate values of $p_{C}$ (see Fig. \ref{fig:Ecological}).

\section*{Acknowledgments}
A. M. thanks Farhan Abedin and Gy\"{o}rgy Szab\'{o} for helpful discussions. A. M. and C. H. acknowledge financial support from the Natural Sciences and Engineering Research Council of Canada (NSERC) and C. H. from the Foundational Questions in Evolutionary Biology Fund (FQEB), grant RFP-12-10.

\bibliographystyle{unsrtnat}

\end{document}